\documentclass[fleqn,usenatbib]{mnras}

\usepackage{newtxtext,newtxmath}

\usepackage[T1]{fontenc}

\DeclareRobustCommand{\VAN}[3]{#2}
\let\VANthebibliography\thebibliography
\def\thebibliography{\DeclareRobustCommand{\VAN}[3]{##3}\VANthebibliography}

\usepackage{graphicx}	
\usepackage{amsmath}	

\usepackage{pdflscape}
\usepackage{longtable}
\usepackage{textcomp}
\usepackage{gensymb}
\usepackage{changepage} 
\usepackage{booktabs}
\usepackage{caption}
\usepackage{siunitx, array, multirow}
\usepackage{subcaption}
\usepackage{ragged2e}

\usepackage{anyfontsize}

\usepackage{lineno}

\usepackage{color,soul}

\title[Probing the Infrared/Radio correlation of the IRAS RBGS]{Probing the Infrared/Radio correlation of the full IRAS Revised Bright Galaxy Sample with MeerKAT and the VLA}

\author[M.E. Moloko et al.]{
Malebo E. Moloko,$^{1,8}$\thanks{E-mail: maleboella@gmail.com}
L. Marchetti,$^{1,2,8}$
T.H. Jarrett,$^{1,7}$
J.J. Condon,$^{3}$
W.D. Cotton, $^{4,6}$
A.M. Matthews,$^{5}$%
\newauthor
T. Mauch,$^{6}$
M. Vaccari$^{8,9,2}$
\\
$^{1}$Department of Astronomy, University of Cape Town, 7701 Rondebosch, Cape Town, South Africa\\
$^{2}$INAF $-$ Istituto di Radioastronomia, via Gobetti 101, 40129 Bologna, Italy\\
$^{3}$Unaffiliated, 2571 Old Lynchburg Road, North Garden, VA 22959, USA\\
$^{4}$National Radio Astronomy Observatory, 520 Edgemont Road, Charlottesville, VA 22903, USA\\
$^{5}$Observatories of the Carnegie Institution for Science, 813 Santa Barbara St, Pasadena, CA 91101, USA \\
$^{6}$South African Radio Astronomy Observatory (SARAO), 2 Fir Street, Black River Park, Observatory, 7925, South Africa\\
$^{7}$Institute for Astronomy, University of Hawaii at Hilo, 640 N Aohoku Pl 209, Hilo, HI 96720, USA\\
$^{8}$Inter-University Institute for Data Intensive Astronomy (IDIA), Department of Astronomy, University of Cape Town, 7701 Rondebosch, Cape Town, South Africa\\
$^{9}$Inter-University Institute for Data Intensive Astronomy, Department of Physics and Astronomy, University of the Western Cape, 7535 Bellville, Cape Town, South Africa
}

\date{Accepted XXX. Received YYY; in original form ZZZ}

\pubyear{2025}

\begin{document}
\label{firstpage}
\pagerange{\pageref{firstpage}--\pageref{lastpage}}
\maketitle

\begin{abstract}
We study the infrared/radio correlation of galaxies in the IRAS Revised Bright Galaxy Sample using new MeerKAT observations at $\rm\nu = 1.28\, GHz$, complemented with VLA data. We classify the objects by primary energy source (Active Galactic Nuclei vs. Star-Forming) and take into account their merger status. With this, we aim to explore the effect of galaxy-galaxy interaction on the total-infrared (TIR)/radio correlation ($q_\mathrm{TIR}$) of star-forming galaxies by comparing the $q_\mathrm{TIR}$ distribution between isolated and interacting/merging sources. We found the median $q_\mathrm{TIR}$ to be $2.61 \pm 0.01$ (scatter = 0.16) for isolated galaxies and $2.51 \pm 0.08$ (scatter = 0.26) for interacting/merging galaxies. Our analysis reveals that interacting/merging galaxies exhibit lower $q_\mathrm{TIR}$ and higher dispersion compared to isolated galaxies, and the difference is marginally significant. Interacting/merging galaxies have redder $W2-W3$ colours, higher star formation rates (SFR) and specific SFR compared to isolated objects. We observe a significant decrease in $q_\mathrm{TIR}$ with increasing radio luminosity for isolated galaxies. Additionally, we find the median ratio of TIR ($8 \,\mu m < \lambda < 1000\, \mu m$) to far-infrared (FIR; $40 \,\mu m < \lambda < 120\, \mu m$) luminosities to be $\left<L_\mathrm{TIR}/L_\mathrm{FIR}\right>\approx2.29$. By examining the relation between $L_\mathrm{TIR}$ and the mid-infrared (MIR) star-formation rate indicator ($L_\mathrm{12\,\mu m}$) employed for our interacting/merging sample, we note a strong and consistent (similar non-linear behaviour) relationship between the TIR/radio and TIR/MIR ratios. Finally, we show that already at $z<0.1$, $q_\mathrm{TIR}$ exhibits a dependence on stellar mass, with more massive galaxies displaying a lower $q_\mathrm{TIR}$.
\end{abstract}




\section{Introduction}\label{intro}
Observations with the Infrared Astronomical Satellite (IRAS; \citealt{Neugebauer1984}) were pivotal, as they not only provided the first unbiased view of the infrared sky, but also revealed a plethora of high-luminosity infrared galaxies/sources. Some of the most powerful infrared galaxies are known as luminous and ultra-luminous infrared galaxies (i.e., LIRGs; $10^{11} \, {L_\mathrm{\odot}} \leq L_\mathrm{TIR} < 10^{12}\, {L_\mathrm{\odot}} $ and ULIRGs; $10^{12} {L_\mathrm{\odot}} \leq L_\mathrm{TIR}\leq 10^{13} \, {L_\mathrm{\odot}} $). These systems rank among the most intensely star-forming galaxies in the universe. Although rare in the local universe, U/LIRGs account for more than half of the cosmic infrared (IR) background and dominate the star formation rate at $z \gtrsim 1$ \citep{LeFloch2005ApJ...632..169L, Gruppioni2013MNRAS, Magnelli2013, Hung2014ApJ...791...63H, song2022ApJ...940...52S}. Additionally, they are often associated with interacting/merging galaxies (e.g., \citealt{Sanders1996}), which play a significant role in triggering intense nuclear starbursts or fuelling powerful Active Galactic Nuclei (AGN). Thus, studying these populations is crucial for our understanding of processes that drive galaxy growth across the universe.

The extreme luminosities observed in U/LIRGs are thought to be powered by star formation (SF), accreting super-massive black holes (SMBHs; i.e., AGNs) or a combination of both \citep{Sanders1996}. Although star formation dominates far-infrared (FIR) emission, AGNs can also significantly contribute to and even dominate the emission of ULIRGs at near and mid-infrared wavelengths (e.g., \citealt{Farrah2003MNRAS.343..585F, Vega2008A&A...484..631V,  Papaefthymiou2022}). Understanding the individual contributions of SF and AGNs in these systems is challenging, as the luminosity originates from nuclear regions heavily attenuated by dust. However, a multi-wavelength approach allows us to circumvent this challenge. Infrared emission traces the dust heated through star formation, providing insight into ongoing star formation activity. Meanwhile, radio observations can penetrate through the dust and gas that obscure these objects at other wavelengths. Together, the synergy of infrared and radio wavelengths offers a valuable opportunity to investigate these galaxies.

Star-forming galaxies exhibit a tight relation between their infrared and radio flux densities, defining what is known as the "Infrared/radio correlation" (\citealt{dejong1985, Helou1985, condon1992, yun2001, bell2003}), which holds over four orders of magnitude in luminosity with a low scatter of about 0.26 dex \citep{yun2001}. Studies of the nearby universe (e.g., \citealt{Helou1985}) have shown that not only does this correlation hold for the late-type galaxies ranging from normal galaxies to the most violent starbursts in ULIRGs (e.g., see review by \citealt{Sanders1996} and references therein), but also for early-type galaxies with low levels of star formation and interacting systems of mixed morphological composition (e.g., \citealt{Domingue2005, Donevski2015MNRAS.453..638D}). Due to its stability, consistency, and tightness, this correlation has found many applications in astrophysics. For instance, it has been used to identify radio-excess sources in distant star-forming galaxies, uncovering the presence of hidden and highly obscured active galactic nuclei (AGN; e.g., \citealt{delmoro2013}). Additionally, researchers such as \citet{Carilli1999, Dunne2000} used FIR/radio flux ratios measured in the observer's frame to estimate the redshifts of high-redshift submillimeter galaxies.
 
After the first indications of the infrared/radio correlation ($q$) by \cite{vanderKruit1973} using ground-based observations at $\rm 10 \,\mu$m and $\rm 1.4$ GHz, the ubiquity and tightness of this correlation only became widely recognised through observations with the Very Large Array (VLA) and IRAS which measured far-infrared properties of many galaxies in the local ($z < 0.15$) universe (e.g., \citealt{Werner2004, Pilbratt2010, Murakami2007}). The birth and death of massive ($M\,> 8 \, \rm{M_\mathrm{\odot}} $) stars are key factors shaping the observed relation between infrared and radio emission, as they have a profound effect on their surroundings. When the UV/optical radiation from these young massive stars heats the dust within their surrounding giant birth clouds, the dust grains absorb and reradiate the emission in the infrared regime. On the other hand, when these very stars die as supernovae (SNe), they accelerate cosmic-ray electrons into the galaxy's magnetic field, thereby contributing to nonthermal radio continuum emission \citep{Condon1991}.

Historically, the FIR diagnostic has been considered reliable for assessing star formation activity because dust heated by starbursts emits predominantly in the far-IR range, while dust heated by AGN in the compact torus around the nucleus emits mainly at shorter mid-IR wavelengths. However, distinguishing between AGN-related components and the FIR continuum is challenging due to potential contamination from various sources, including cool dust components in the torus and AGN illumination of larger-scale regions. In the absence of FIR information, mid-infrared (MIR) measurements from sensitive instruments such as WISE can be employed as proxies to estimate the total IR (TIR) luminosity (e.g., \citealt{cluver2017}), thus also facilitating exploration of mid-IR/radio correlations (e.g., \citealt{yao2022}). Generally, the infrared/radio correlation can serve as a helpful tool for identifying AGNs when there is a lack of good-quality nuclear spectra and high-resolution IR data. 

The strong correlation between infrared and radio emission allows the radio continuum to serve as an unbiased tracer of star formation, a capability further enhanced by the advent of facilities such as MeerKAT, ASKAP, and LOFAR, which reach lower flux densities with unprecedented sensitivity and resolution. Additionally, radio emission offers several advantages over other SFR tracers, making it a powerful tool for obtaining a comprehensive view of the cosmic star formation history up to very high redshifts (e.g., \citealt{Madau2014, Novak2017, Delhaize2017, Matthews2021}). 

In this paper, we aim to measure the infrared-radio correlation of galaxies in the IRAS Revised Bright Galaxy Sample (RBGS; \citealt{Sanders2003}), a nearby sample of dusty star-forming galaxies, using data from the MeerKAT Luminous Infrared Galaxies (MeerLIRGs) project, a recent observing programme conducted with MeerKAT (\citealt{Condon2021}) complemented by VLA archival data. The sample includes galaxies with diverse spectral types—classified using conventional optical-line ratio diagrams (i.e., Seyferts, LINERs, and Starbursts)—and spans different interaction stages, making it ideal for studying the processes driving extreme infrared emission. 

\begin{figure*}
    \centering
    \includegraphics[width=\textwidth]{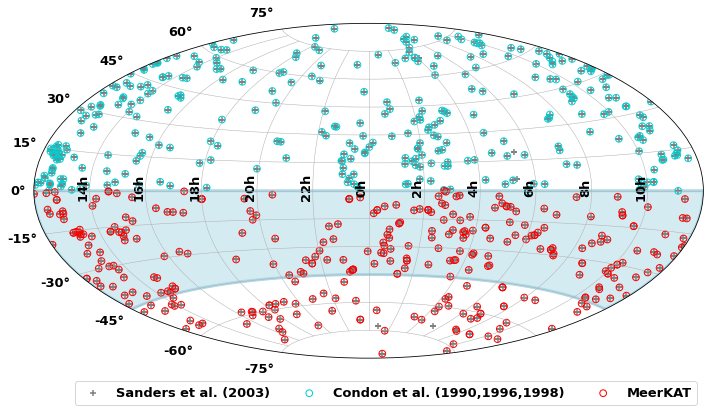}
    \caption{Aitoff sky coverage showing the 629 RBGS galaxies \citep{Sanders2003} as grey crosses. MeerKAT observed all southern RBGS galaxies except the Milky Way satellites LMC and SMC \citep{Condon2021}. Galaxies north of $\rm \delta = -45 \deg$ were observed by various configurations of the VLA \citep{Condon1990, Condon1996, Condon1998}. Despite the observed overlap, as indicated by the highlighted region, only MeerKAT flux densities are employed for analysing all sources in the southern hemisphere.}
    \label{multiwavelength-sky-coverage}
\end{figure*}

We use literature spectroscopic classifications and WISE colour-colour diagnostics to categorise the sources based on their dominant activity (AGN vs. star formation) and investigate their properties. We first analyse the full sample, also referred to as "Total Sample", without distinguishing between isolated and interacting/merging systems. This approach allows us to create, for the first time, a nearly complete radio counterpart catalogue of the all-sky RBGS (see Figure \ref{multiwavelength-sky-coverage}). Both isolated and interacting/merging systems are initially treated as single entities, as in \citet{Sanders2003}, though higher-resolution observations reveal that some consist of multiple interacting/merging components. However, doing this enables direct comparison with historical IRAS-based studies and allows us to quantify how mergers might influence the results. Next, we focus on sources that are truly isolated (i.e., not unresolved merging systems in \citealt{Sanders2003}), referred to as the "Isolated Sample", allowing us to uncover intrinsic trends. Lastly, we analyse the systems that MeerKAT and WISE have resolved into individual components (the "Interacting/Merging Sample"), enabling a detailed study of their infrared and radio properties.

This paper is organised into 5 Sections. In Section \ref{sample_data_discription}, we detail our process for selecting the sample, provide information about the data, and outline the morphological classifications, redshifts and luminosities of the sources. Section \ref{results} is dedicated to presenting our findings, including a comparison with the literature. In Section \ref{discussion}, we discuss the limitations and implications of our results. Finally, Section \ref{conclusion} summarises major results and conclusions. Throughout this paper, we adopt a $\Lambda$CDM cosmology with $ H_\mathrm{0} = \mathrm{70 \, km\, s^{-1} \,Mpc^{-1}}$, $\Omega _\mathrm{m}=0.3$ and $ \Omega _\mathrm{\Lambda}=0.7$, and use equations from \cite{Condon2018PASP..130g3001C} to calculate galaxy distances and intrinsic source properties.  All magnitudes are in the Vega system unless stated otherwise.

\section{Sample and Data}\label{sample_data_discription}
This Section provides details about the sample selection and the data used in our analysis. We also explain how we obtained the redshift measurements of the objects and classified them into different categories. Lastly,  we describe how the radio spectral luminosities and $q$ parameters were computed.

\subsection{Sample Selection}
The objects we use in this sample are drawn from the IRAS Revised Bright Galaxy Sample (RBGS; \citealt{Sanders2003}), which comprises 629 individual galaxies and interacting/merging systems in the extragalactic sky defined by galactic latitudes |b| > 5 $\deg$. These objects have IRAS flux densities greater than 5.24 Jy at $ \rm \lambda = 60 \, \mu m $. Further details about the original catalogue are provided in Section \ref{sec:IRdata}. In brief, $\rm 300$ of the RBGS sources are located in the southern hemisphere, while the remaining $\rm 329$ are in the northern hemisphere. In recent years, we have followed the southern IRAS sources for observations with MeerKAT at $\rm 1.28$ GHz \citep{Condon2021} through two observing campaigns: Director’s Discretionary Time (DDT-20200520-TM-01) and "Open Time" (SCI20210212-TJ-01). More details on these observations are provided in Section \ref{sec:radiodata}, but for full information, the reader is referred to \cite{Condon2021}. We use this as the primary radio data for our analysis. All but two southern RBGS objects, namely the Large and Small Magellanic Clouds (LMC and SMC), were observed with MeerKAT. \cite{Condon2021} excluded these Magellanic Clouds from the sample because they are considered satellites of the Milky Way, making them not representative of the galaxy population due to a location bias. Thus, the remaining sample contains 298 sources, and we dub it the  "southern" sample.

In addition to the southern sample, complementary data observed with the VLA (e.g., \citealt{Condon1990, Condon1996}) for the RBGS objects located in the Northern Hemisphere (see Section \ref{sec:radiodata} for details) are available. We utilised these data to complement our MeerKAT study, allowing us to present, for the first time, the radio analysis of the (almost) full RBGS sample (Figure \ref{multiwavelength-sky-coverage}). All but three of these northern targets were detected by the VLA studies considered in this work, namely IC $\rm 0356$, IRAS F$\rm 05170+0535$ and IRAS $\rm 05223+1908$. IC 0356 was observed with the VLA by \cite{Condon1987}, and while it was reported to have a flux density $S = 29.5 \,\rm{mJy}$, the source was not clearly resolved from its nearby companions, and therefore the radio flux was not deemed accurate. IRAS F05170+0535 is the T Tauri star HD 34700 \citep{Sterzik2005A&A...434..671S} and IRAS 05223+1908 is a young stellar object \citep{Chu2017ApJS..229...25C}, and therefore neither object should be in the RBGS sample in the first place (see also \citealt{Reynolds2022A&A...664A.158R}). The remaining sample is left with $\rm 326$ sources, and we dub it the "northern" sample. 

\begin{figure}
    \centering
    \includegraphics[width=0.48\textwidth]{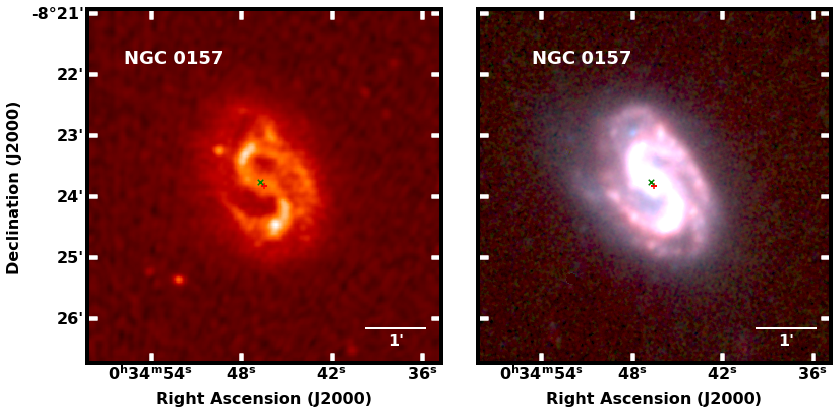}\\
    \hspace{0.4cm}
    \includegraphics[width=0.48\textwidth]{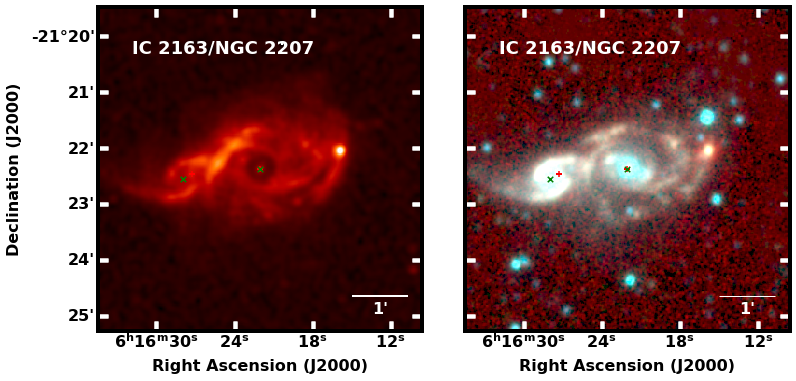}\\
    \hspace{0.4cm}
    \includegraphics[width=0.48\textwidth]{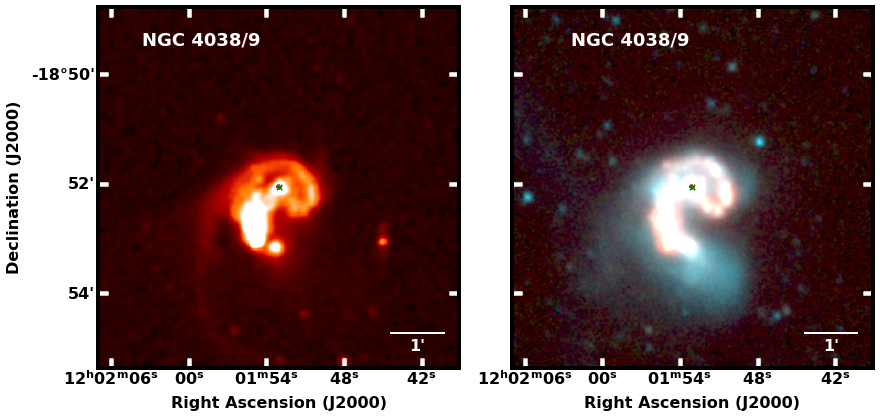}
    \caption{MeerKAT (left) and WISE 3-colour (right) images of some of the sources in our sample depicting the different interaction stages found in this sample. The top row shows an example of an isolated galaxy (NGC $\rm 0157$), while the middle row depicts an interacting galaxy (IC $\rm 2163$/ NGC $\rm 2207$) and the last row shows a merging galaxy (NGC $\rm 4038/9; \, aka\, Arp\, 244$). The positions of the source are also overlaid on the images, identified as red and green crosses for MeerKAT and WISE measurements, respectively.}
    \label{example|-images}
\end{figure}

Since we study the full-sky RBGS sample in this work, our sample also includes the 201 sources targeted by the Great Observatories All-Sky LIRG Survey (GOALS; \citealt{Armus2009}), a multi-wavelength survey of LIRGs in the nearby universe ($z<0.088$). However, while the GOALS sources are part of our sample, they are not directly highlighted because we do not make use of the additional (i.e. Spitzer) data available for them, which will be subject to future work. In general, the RBGS sample spans a wide range of spectral types (Starbursts, AGNs, and LINERs; see Section \ref{nuclear_class}) and interaction stages (isolated, interacting, and merging systems). In Figure \ref{example|-images}, we show examples of sources at these different stages, such as NGC $\rm 0157$ (isolated), IC $\rm 2163$/ NGC $\rm 2207$ (interacting) and NGC $\rm 4038/9$ (merging) found in our sample. 

\subsection{Radio: MeerKAT and VLA}\label{sec:radiodata}

The primary radio data used in this paper are the L-band observations taken with the MeerKAT array at $\rm \nu = 1.28$ GHz by \cite{Condon2021}. The observations were achieved over several observing runs in May $\rm 2020$ and February $\rm 2021$ using $\rm 60\,$ to $\rm 64$ operating antennas. The observations were carried out for all $\rm 298$ RBGS sources located in the southern hemisphere ($\rm J2000 \, \delta \,< \,0 \degr$). Each observing session was $\rm \sim 8$ hours long, and each source was observed with five 3-minute snapshots. The $\rm 856$ to $\rm 1712\, MHz$ frequency range was split into $\rm 4096$ spectral channels to reduce bandwidth smearing. The integration time was $\rm 8$ seconds to reduce time smearing. For a full description of the observations, imaging and reductions, the reader is referred to \cite{Condon2021}. In brief, the initial data reductions were conducted using the $Obit$ software \citep{cotton2008obit} to flag bad data, remove strong radio frequency interference (RFI), calibrate the data and combine the pointings into a single image. The resulting images have $\rm \approx 7\farcs5$ FWHM resolution and rms fluctuations $\rm \sigma_{n}  \approx \, 20\, \mu Jy\, beam^{-1}$. 

In addition to our MeerKAT observations in the south, $\rm 1.49$ and $\rm 1.425 \, \mathrm{GHz}$ observations of nearly all RBGS sources with $\delta > -45 \degree$ were made with several configurations of the VLA (at $\rm 1.49$ GHz by \citealt{Condon1987, Condon1990} and at $\rm 1.425$ GHz by \citealt{Condon1996}). Almost all have $1.4 \, \mathrm{GHz}$ flux densities $\gtrsim$  $\rm 30 \, \mathrm{mJy}$ and surface brightness temperatures $T_\mathrm{b} \gtrsim 0.8 \, \mathrm{K}$, equivalent to $1.28\, \mathrm{GHz}$ $T_\mathrm{b}\, \gtrsim 1.2 \,\mathrm{K}$ for sources with typical spectral indices $\rm \alpha \approx -0.7$. Not all the RBGS objects in the northern hemisphere were observed in these targeted observations. Hence, for those objects without measurements, we employed data from $\rm 1.4 \, GHz$ NRAO VLA Sky Survey (NVSS; \citealt{Condon1998}) made at comparable depth. We combined all the measurements of the sources from these catalogues to create a comprehensive radio catalogue of the northern sources to be used along with the MeerKAT measurements for the southern sample.

\begin{figure}
    \centering
    \includegraphics[width=0.49\textwidth]{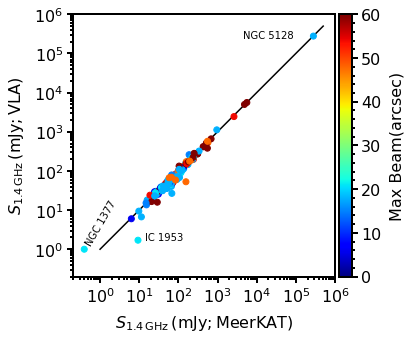}
    \caption{Comparison of the VLA $\rm 1.49$ and $\rm 1.425$ GHz (\citealt{Condon1990, Condon1996}, respectively) maximum-beam flux densities (y-axis) with MeerKAT 1.28 GHz (\citealt{Condon2021}) flux densities (x-axis) for objects with measurements in both catalogues. All the flux densities are converted to $\rm 1.4 \, GHz$ assuming a spectral index of $\rm - 0.7$. }
    \label{radio_flux_comps}
\end{figure}

We converted all the radio flux densities to $ \nu = 1.4$ GHz assuming a spectral index $\mathrm{\alpha} \equiv d \mathrm{ln}(S)/ d \mathrm{ln}(\nu) = -0.7$ (typical for most SFGs) to compare them with the literature. $\rm 194$ of the $\rm 629$ IRAS RBGS sources have both MeerKAT and VLA measurements, as indicated by the highlighted region in Figure \ref{multiwavelength-sky-coverage}. The majority of the sources with both VLA (\citealt{Condon1990}, \citealt{Condon1996}) and MeerKAT (\citealt{Condon2021}) pointed observations lie close to the one-to-one line in the plot depicting the VLA flux densities in the maximum beam (i.e., the largest beam of each observation) against the MeerKAT flux densities (see Figure \ref{radio_flux_comps}) once converted to the same reference frequency. This indicates that the flux densities are comparable, with most sources agreeing within approximately $ \pm 23\%$) and can therefore be used concurrently.

Among the various VLA flux density measurements available for these sources, we chose to consider those with the lowest resolution (as indicated by the colour bar in Figure \ref{radio_flux_comps}) because they provide a better representation of the sources' extent and yield more reliable flux density estimates consistent with the MeerKAT measurements. In addition, Figure 4 of \cite{Condon2021} already shows that the NVSS and MeerKAT measurements are comparable, with most RBGS objects lying near the expected ratio ($\left<\mathrm{S(1.4\, GHz)}/S\mathrm{(1.28\, GHz)}\right> \approx 0.94$). As a result, we are confident that the use of the NVSS fluxes for those few sources lacking pointed VLA observations will be consistent with the rest of the data used in our analysis. 

The total flux densities at $\rm 1.28$ GHz, as reported in \cite{Condon2021}, were estimated for each source by integrating radio emission across its map. The typical rms intensity-proportional error (i.e., $\sigma  / S$) was estimated using a statistical comparison of these MeerKAT flux densities with NVSS flux densities, finding that MeerKAT flux densities are better than NVSS flux densities (see \citealt{Condon2021}), so, for those objects observed by both, we use the MeerKAT flux as reference. The errors associated with the MeerKAT flux densities are not greater than $\sigma /S = 0.05$. Given that the rms noise in our MeerKAT images is only about $0.02\, \rm{mJy/beam}$, significantly lower than the typical flux densities of RBGS sources, noise is a negligible source of error for all but one or two of the weakest radio sources. Similarly, the rms confusion error in VLA total flux densities at $\rm 1.49$ GHz (\citealt{Condon1990}) and $\rm 1.425$ GHz (\citealt{Condon1996}) is smaller than the noise error in their radio maps. With intensity-proportional calibration errors remaining low at $\rm 3-5 \%$ (\citealt{Condon1990,Condon1996}), they are also negligible.

\begin{table*}
\captionsetup{justification=centering}
\caption{The IRAS RBGS Sample Properties with MeerKAT and the VLA. The complete table is available on Zenodo at \url{https://doi.org/10.5281/zenodo.16889781}.}
\centering
\begin{tabular}{lc@{\hspace{0.5\tabcolsep}}cccccccl}
\toprule
\midrule
 Name & RA (J2000) & DEC (J2000)  & $z_\mathrm{helio}$  & $z_\mathrm{cmb}$ & $D_\mathrm{L}$ & log($ L_\mathrm{TIR})$ & $L_\mathrm{1.4\,GHz}$ & $q_\mathrm{TIR}$ & Notes$^*$ \\
   & $\rm hh:mm:ss$ & $\degr:\farcm:\farcs$ &  & & $\mathrm{Mpc}$ & $\mathrm{L_\mathrm{\odot}}$ & $\mathrm{W\,Hz^{-1}}$ & &\\
   
    (1) &  (2) &  (3)  & (4) & (5) & (6) & (7) & (8) & (9)  & (10) \\
\hline
UGC 12914/5 & 00 01 40.7 & +23 29 37 & 0.0145(3) & 0.01334 & 57.80 & 10.89 & 4.72e+22 & 2.23 & S?[1]  \\ 
NGC 0023 & 00 09 55.1 & +25 55 37 & 0.01316(2) & 0.01316 & 57.02 & 11.00 & 2.62e+22 & 2.59 & SB(s)a; Starburst[0]  \\ 
NGC 0034 & 00 11 06.6 & -12 06 27 & 0.01948(0) & 0.01835 & 79.81 & 11.45 & 6.98e+22 & 2.61 & pec; Sy 2[0]  \\ 
NGC 0055 & 00 14 54.5 & -39 11 19 & 0.00051(1) & 0.00051 & 2.19 & 9.06 & 3.16e+20 & 2.57 & SB(s)m?[0]  \\ 
MCG -02-01-051/2 & 00 18 51.4 & -10 22 33 & 0.0271(0) & 0.02597 & 113.59 & 11.46 & 6.62e+22 & 2.64 & [2]  \\ 
NGC 0134 & 00 30 21.6 & -33 14 38 & 0.00479(2) & 0.00479 & 20.62 & 10.59 & 1.08e+22 & 2.57 & SAB(s)bc[0]  \\ 
ESO 079-G003 & 00 32 02.0 & -64 15 14 & 0.00873(0) & 0.00836 & 36.09 & 10.52 & 9.01e+21 & 2.58 & SBb?[0]  \\ 
NGC 0150 & 00 34 16.2 & -27 48 10 & 0.00591(2) & 0.00591 & 25.46 & 10.34 & 4.23e+21 & 2.72 & SB(rs)b?[0]  \\ 
NGC 0157 & 00 34 46.0 & -08 23 53 & 0.00301(2) & 0.00301 & 12.94 & 10.06 & 3.63e+21 & 2.51 & SAB(rs)bc[0]  \\ 
ESO 350-IG038 & 00 36 52.0 & -33 33 19 & 0.0206(0) & 0.01973 & 85.90 & 11.25 & 2.51e+22 & 2.86 & pair[0]  \\ 
NGC 0174 & 00 36 58.4 & -29 28 45 & 0.01177(0) & 0.01085 & 46.92 & 10.88 & 1.17e+22 & 2.82 & SB0/a?(rs)[0]  \\ 
NGC 0224 & 00 42 44.8 & +41 16 13 & 0.0002(1) & 0.0002 & 0.86 & 9.47 & 7.51e+20 & 2.60 & SA(s)b[0]  \\ 
NGC 0232 & 00 42 46.5 & -23 33 31 & 0.02074(2) & 0.02074 & 90.37 & 11.40 & 5.92e+22 & 2.63 & SB(r)a? pec[0]  \\ 
NGC 0247 & 00 47 06.0 & -20 45 48 & 0.00091(1) & 0.00091 & 3.90 & 8.65 & 3.42e+19 & 3.12 & SAB(s)d[0]  \\ 
NGC 0253 & 00 47 33.1 & -25 17 15 & 0.00086(1) & 0.00086 & 3.69 & 10.59 & 9.13e+21 & 2.64 & SAB(s)c[0]  \\ 
NGC 0278 & 00 52 04.3 & +47 33 01 & 0.00209(0) & 0.00123 & 5.28 & 9.36 & 4.65e+20 & 2.70 & SAB(rs)b[0]  \\ 
NGC 0289 & 00 52 42.8 & -31 12 19 & 0.0058(2) & 0.0058 & 24.99 & 10.15 & 3.62e+21 & 2.60 & SB(rs)bc[0]  \\ 
MCG +12-02-001 & 00 54 04.0 & +73 05 13 & 0.01042(2) & 0.01042 & 45.05 & 11.12 & 3.03e+22 & 2.65 & [1]  \\ 
UGC 00556 & 00 54 49.9 & +29 14 42 & 0.02376(2) & 0.02376 & 103.76 & 11.29 & 5.71e+22 & 2.54 & S?[0]  \\ 
NGC 0300 & 00 54 52.9 & -37 41 09 & 0.00048(1) & 0.00048 & 2.06 & 8.42 & 4.96e+19 & 2.73 & SA(s)d[0]  \\ 
NGC 0317B & 00 57 40.9 & +43 47 37 & 0.01811(0) & 0.01721 & 74.79 & 11.15 & 4.37e+22 & 2.52 & SB?[2]  \\ 
NGC 0337 & 00 59 49.6 & -07 34 52 & 0.00479(2) & 0.00479 & 20.62 & 10.09 & 5.44e+21 & 2.36 & SB(s)d[0]  \\ 
IC 1623A/B & 01 07 46.3 & -17 30 32 & 0.0203(0) & 0.01933 & 84.14 & 11.69 & 2.02e+23 & 2.40 & [1]  \\ 
MCG -03-04-014 & 01 10 08.5 & -16 51 14 & 0.03349(0) & 0.03251 & 142.89 & 11.64 & 1.09e+23 & 2.61 & [0]  \\ 
ESO 244-G012 & 01 18 08.6 & -44 27 40 & 0.02093(0) & 0.02029 & 88.38 & 11.35 & 5.57e+22 & 2.62 & [2]  \\ 
NGC 0470 & 01 19 45.0 & +03 24 36 & 0.00991(2) & 0.00991 & 42.83 & 10.64 & 8.94e+21 & 2.70 & SA(rs)b[0]  \\ 
CGCG 436-030 & 01 20 01.4 & +14 21 35 & 0.03104(0) & 0.02999 & 131.57 & 11.67 & 1.05e+23 & 2.66 & ?[2]  \\ 
UGC 00903 & 01 21 47.4 & +17 35 33 & 0.01101(2) & 0.01101 & 47.62 & 10.75 & 1.15e+22 & 2.70 & S?[0]  \\ 
NGC 0520 & 01 24 34.4 & +03 47 29 & 0.00761(0) & 0.00657 & 28.32 & 10.85 & 1.66e+22 & 2.64 & pec[1]  \\ 
NGC 0598 & 01 33 54.0 & +30 40 07 & 0.00021(1) & 0.00021 & 0.90 & 9.13 & 3.33e+20 & 2.62 & SA(s)cd[0]  \\ 
NGC 0613 & 01 34 17.8 & -29 25 10 & 0.00494(0) & 0.00414 & 17.81 & 10.52 & 8.94e+21 & 2.58 & SB(rs)bc; Sy ?[0]  \\ 
ESO 353-G020 & 01 34 49.4 & -36 08 25 & 0.01592(0) & 0.0152 & 65.95 & 11.02 & 5.06e+22 & 2.33 & S$0^+$?[0]  \\ 
NGC 0625 & 01 35 05.8 & -41 26 17 & 0.00088(2) & 0.00088 & 3.78 & 8.43 & 1.94e+19 & 3.15 & SB(s)m?[0]  \\ 
ESO 297-G011/012 & 01 36 24.7 & -37 19 56 & 0.0173(0) & 0.01661 & 72.15 & 11.13 & 2.91e+22 & 2.67 & [2]  \\ 
NGC 0628 & 01 36 41.2 & +15 47 29 & 0.00234(1) & 0.00234 & 10.05 & 9.95 & 2.26e+21 & 2.61 & SA(s)c[0]  \\ 
IRAS F01364-1042 & 01 38 52.6 & -10 27 15 & 0.04841(0) & 0.04747 & 210.91 & 11.82 & 7.84e+22 & 2.93 & LINER[0]  \\ 
NGC 0643B & 01 39 12.9 & -75 00 40 & 0.01338(0) & 0.01327 & 57.50 & 10.88 & 1.73e+22 & 2.66 & SB0?[0]  \\ 
NGC 0660 & 01 43 02.1 & +13 38 45 & 0.00338(2) & 0.00338 & 14.54 & 10.63 & 1.02e+22 & 2.63 & SB(s)a pec; Sy 2[0]  \\ 
III Zw 035 & 01 44 30.0 & +17 06 04 & 0.02772(0) & 0.02675 & 117.08 & 11.62 & 6.97e+22 & 2.78 & [0]  \\ 
NGC 0693 & 01 50 30.9 & +06 08 43 & 0.00457(2) & 0.00457 & 19.67 & 9.89 & 2.18e+21 & 2.56 & S0/a?[0]  \\ 
NGC 0701 & 01 51 04.5 & -09 42 04 & 0.0051(2) & 0.0051 & 21.96 & 10.06 & 2.74e+21 & 2.63 & SB(rs)c[0]  \\ 
NGC 0695 & 01 51 14.0 & +22 34 55 & 0.03071(0) & 0.03071 & 134.80 & 11.66 & 1.48e+23 & 2.49 & S0? pec[1]  \\ 
NGC 0697 & 01 51 17.5 & +22 21 30 & 0.00811(2) & 0.00811 & 35.00 & 10.50 & 6.23e+21 & 2.71 & SAB(r)c?[0]  \\ 
NGC 0716 & 01 52 58.8 & +12 42 25 & 0.01481(2) & 0.01481 & 64.24 & 10.96 & 2.01e+22 & 2.67 & SBa?[0]  \\ 
UGC 01385 & 01 54 53.1 & +36 54 59 & 0.01475(2) & 0.01475 & 63.98 & 10.85 & 9.31e+21 & 2.89 & SB0/a[0]  \\ 
UGC 01451 & 01 58 30.2 & +25 21 36 & 0.01185(2) & 0.01185 & 51.29 & 10.73 & 1.25e+22 & 2.65 & SB?[0]  \\ 
NGC 0772 & 01 59 19.5 & +19 00 23 & 0.00825(0) & 0.00733 & 31.61 & 10.54 & 8.86e+21 & 2.60 & SA(s)b[0]  \\ 
IRAS 01583+6807 & 02 02 17.4 & +68 21 41 & 0.01193(0) & 0.01146 & 49.59 & 10.72 & 2.76e+22 & 2.29 & [0]  \\ 
NGC 0835 & 02 09 23.8 & -10 08 13 & 0.01359(0) & 0.01275 & 55.22 & 10.77 & 1.72e+22 & 2.54 & SAB(r)ab? pec; Sy 2[0]  \\ 
NGC 0838 & 02 09 38.8 & -10 08 46 & 0.01284(0) & 0.01201 & 51.99 & 11.02 & 2.94e+22 & 2.56 & SA$0^0$(rs) pec?[0]  \\ 
\bottomrule

\multicolumn{10}{p{\linewidth}}{NOTE:- The column descriptions are (1) The Common optical counterpart names to the galaxy systems. (2)-(3) The IRAS positions of the sources in J2000, taken from from \citet{Sanders2003}. (4) The heliocentric redshift of the sources, with the number in parentheses showing the flag indicating whether the distance is based on the non-relativistic redshift reported by NED (0), or the NED redshift-independent distance (1) and the CosmicFlowsIII redshift-independent distance (2). (5) Is the redshift corrected to the CMB reference frame, while (6) is the corresponding luminosity distance assuming $H_\mathrm{{0}} =70\, \mathrm{km\,  s^{-1} \, Mpc^{-1}}$. (7) The total infrared luminosity in units of $L_\mathrm{\odot}= 3.83 \times 10^{26} \, \mathrm{W}$ of the entire system from \citet{Sanders2003} scaled to our cosmology and distances. (8) The radio spectral luminosity of the objects converted to 1.4 GHz from the MeerKAT ($\rm 1.28$ GHz) and the VLA ($\rm 1.49,\,1.425$ and $\rm 1.4$ GHz). (9) Is the corresponding infrared/radio correlation ($q_\mathrm{TIR}$). The rms error associated with the individual $q_\mathrm{TIR}$ values is less than $0.05$ for most sources, with a tail extending above this threshold for a few cases. NGC 1377 is an example of such an object, which exhibits high uncertainty and anomalous behaviour. (10) Indicates the morphology and activity type of the sources from NED. The number in the last bracket indicates the interaction type of the source/s and the corresponding group in our analysis: 0-isolated, 1-merging, and 2-interacting.}  \\
\end{tabular}
\label{meerligs_props_all}
\end{table*}


\subsection{Far and Mid-infrared Data}\label{sec:IRdata}
The primary infrared data used in this study is from the Infrared Astronomical Satellite \citep[IRAS; ][]{Neugebauer1984} Revised Bright Galaxy Sample (RBGS; \citealt{Sanders2003}). The IRAS  satellite  observed the full sky at wavelengths $ \rm 12,\, 25,\, 60 \, and \,100 \,\mu m$ with angular resolution varying from $\rm \sim 0\farcm5 \,at \, 12\, \mu m$ to $\rm \sim 2' \,at \, 100\, \mu m.$ For full details about the RBGS, refer to \citet{Sanders2003}. We use this, together with our radio data, to compute the infrared/radio correlation and investigate the properties of our "total" and "isolated" samples. In addition to the far-infrared, the objects in our study have mid-infrared measurements obtained from the Wide-field Infrared Survey Explorer (WISE; \citealt{wright2010}). The WISE satellite observed the entire sky in mid-infrared wavelengths using four wavebands centred on $\rm 3.4 \, \mu m$ (W$1$), $\rm 4.6\, \mu m$ (W2), $\rm 12 \, \mu m$ (W3), and $\rm 22\, \mu m$ (W4). These bands have Full Width at Half Maximum (FWHM) resolution drizzled images of $\rm 5\farcs9, \, 6\farcs5, \, 7\farcs0 \, and\, 12\farcs4 $, respectively. 

It is worth mentioning that the WISE survey specifically provided photometry tailored for point sources. Large and resolved systems were often treated as agglomerations of point sources, or measurements would miss significant flux. Hence, any work focusing on nearby extended galaxies requires reprocessing the available WISE imaging and carefully extracting resolved galaxies. For our sample, we utilise the WISE measurements from the Extended Source Catalogue (WXSC; \citealt{Jarrett2013, Jarrett2019, jarrett2023}) curated by the WISE team. The WXSC provides improved imaging and source measurements compared to the published ALLWISE catalogues tailored for point sources. The WXSC database incorporates custom mosaic construction and characterisation, meticulous foreground star (and background galaxy) removal, local background estimation, and a full suite of size/orientation, photometry, surface brightness, and radial profile measurements (see \citealt {Jarrett2019, jarrett2023} for more details).

\subsection{Redshifts}\label{Redshift}
Instead of relying on the redshift measurements reported in \cite{Sanders2003}, we collected updated measurements, where available, for each source. We obtained heliocentric redshift ($ z_\mathrm{helio}$) measurements for all the objects in our sample from numerous sources including the NASA/IPAC Extragalactic Database (NED)\footnote{\url{https://ned.ipac.caltech.edu/}}, the Spitzer Survey of Stellar Structure in Galaxies ($S ^4$G; \citealt{Sheth2010}), the Two Micron All Sky Survey (2MASS; \citealt{Skrutskie2006}) and the 2MASS Extended Source Catalog (XSCz; \citealt{Jarrett2004}). We further corrected the initial $z_\mathrm{helio}$ for all the sources to the cosmic microwave background (CMB) reference frame to be consistent with the literature. Despite this correction, the overall differences between $z_\mathrm{helio}$ and $z_\mathrm{cmb}$ are minimal, typically around $\Delta z \sim 0.001$.

All these redshift measurements are presented in Table \ref{meerligs_props_all}, specifically in the column labelled "$z_\mathrm{helio}$" and "$z_\mathrm{cmb}$". The flag indicating whether the distance (hence luminosity) is based on redshift (0) or NED redshift-independent distance (1), or CosmicFlowsIII \citep{Tully2016} redshift-independent distance (2) is shown in parentheses in the "$z_\mathrm{helio}$" column. The conversions between the CMB-frame redshift and the luminosity distance are performed using the analytic method of \cite{Wickramasingh2010} with $H_{0} = 70 \, \mathrm{km \, s^{-1} \, Mpc^{-1}}$ and $\Omega_\mathrm{m}\, = \,0.3$. Overall, the performed conversions ensure that the distances of these objects are updated. Out of the 624 sources, 593 (267 south and 326 north; $\rm 95\%$) isolated galaxies and merging systems have redshift or distance measurements. For the remaining 31 ($\rm 5\%$) now resolved interacting/merging sources (i.e., by MeerKAT), we employed the redshift/distance of one of the components in the system.

\subsection{Classifications}\label{nuclear_class}
In our analysis, we employ different classification schemes to take into account the different levels of detail one can access for these very well-studied sources. However, we note that the different classifications probe distinct regimes of AGN activity. Literature-based classifications compiled in NED comprise multi-wavelength classes drawn from various diagnostic methods reported in the literature. As such, these classifications represent a heterogeneous set of approaches and can include AGNs with varying contributions to their host galaxies’ bolometric luminosities. In contrast, WISE mid-infrared colour selection primarily identifies AGNs that contribute significantly to the mid-infrared emission (see, e.g., \citealt{lacy2004ApJS..154..166L, Stern2005ApJ...631..163S}). As a result, some level of inconsistency between the two classification schemes is expected. Nevertheless, we present both classifications to highlight the complementary information provided by each approach.

\subsubsection{Literature based classes} 
Most of the RBGS galaxies have extensive ancillary data available, including classifications; therefore, as a first attempt to classify them, we obtained e.g., optical morphology, activity type, and radio properties of the sources from the literature (before our MeerKAT study, the latter would have been available only for a portion of the full RBGS). For this, we relied mostly on the results from NED, and we used this as the reference for the "Total Sample" analysis to provide a more general overview of the RBGS galaxies. The database provided us with morphological classifications for most of the sources. In addition, it has a range of activity type classifications for these objects, including Starburst, HII, $\rm H {\alpha}$ Emission-Line Galaxy, Flat-Spectrum Radio Source, and various Seyferts (Sy) and LINER types based on optical spectroscopy. Some objects also had their radio morphology (e.g., jet, FR I) classifications and the host galaxy morphology already identified. The morphological and activity type classifications of these sources are given in Table \ref{meerligs_props_all} in the column called "Notes". 

In our analysis presented in Section \ref{total_sample}, we consider AGNs to include all Seyferts and those with radio jet/lobe morphologies. We leave LINERs as a distinct class due to the ongoing debate surrounding their nature. To reduce contamination, both the AGNs and LINERs are excluded when we carry out our TIR-radio analysis calculations. We also exclude the interacting/merging systems as their classifications are likely uncertain. Nonetheless, these populations are still presented in all the plots for a comprehensive representation. For a detailed breakdown of the classifications of this sample based on the NED results, refer to Table \ref{classification_breakdown}.

\subsubsection{Separating objects using mid-IR colours}
Mid-infrared colours have been proven to be a useful diagnostic tool for determining the underlying characteristics of a given sample (see \citealt{wright2010, Jarrett2011, Yao2020}). In the WISE colour-colour diagram, the objects are projected in the $\rm W1 - W2$ vs $\rm W2 - W3$ plane (\citealt{wright2010}), where they occupy different regions. Stars and early-type galaxies have colours close to zero, while brown dwarfs are very red in $\rm W1 - W2$, spiral galaxies are red in $\rm W2 - W3$, and ULIRGs exhibit red colours in both. It is important to note that although this diagnostic is valuable for gaining insights into the interplay between AGNs and host galaxies, there is some overlap between different populations. This overlap becomes worse when factors such as obscuration and redshift evolution are considered. Nevertheless, the W1, W2, and W3 WISE bands are useful diagnostics for determining the prevalence of star formation and the relative dominance of a radiative AGN \citep{stern2012}. As a second step in our analysis, we classified our "Isolated" and resolved "interacting/merging" objects into distinct categories based on their WISE MIR diagnostics. This classification is detailed in Section \ref{wise_col}, and it serves as a basis for comparison with the NED classification results mentioned above. See Appendix \ref{wise-ned_class_perfomance} for an evaluation of the comparison between the two schemes. For a comprehensive classification breakdown of this subset, also refer to Table \ref{classification_breakdown}.

\begin{table*}
    \centering
        \begin{tabular}{cccccccc}
    \hline
    & & \multicolumn{2}{c}{South} & \multicolumn{2}{c}{North} & \multicolumn{2}{c}{All Sky} \\
    \cmidrule(lr){3-4} \cmidrule(lr){5-6} \cmidrule(lr){7-8}
    Sample & Class & \# & Fraction & \# & Fraction & \# & Fraction \\
    \hline
    \multirow{3}{*}{Total - Isolated}&AGN & 46 & 15\% & 40 & 12\% & 86 & 14\% \\
    &LINER& 4 & 1\% & 7 & 2\% & 11 & 2\% \\
    &Other & 223 & 75\% & 202 & 62\% & 425 & 68\% \\
    Total - Interacting/Merging & - & 25 & 8\% & 77 & 24\% & 102 & 16\% \\
    \hline
    Total Sample & All & 298 & 100\% & 326 & 100\% & 624 & 100\% \\
    \hline
    \multirow{5}{*}{Isolated}& AGN & 22 & 8\% & 13 & 5\% & 35 & 7\% \\
    &Warm-AGN& 26 & 10\% & 20 & 8\% & 46 & 9\% \\
    &SF& 208& 76\% & 195 & 78\% & 403 & 77\% \\
    &Spheroid& 0 & 0\% & 2& 1\% &2 & 0\% \\
    &Intermediate Disk& 17 & 6\% & 19 & 8\% & 36 & 7\% \\
    &All& 273 & 100\% & 249 & 100\% & 522 & 100\% \\
    \hline
    \multirow{5}{*}{Classified - Interacting/Merging$^{[**]}$}& AGN & 2 & 6\% & - & - & - & -\\
    &Warm-AGN& 2 & 6\%  & - & - & - & -\\
    &SF&  26& 84\%  & - & - & - & - \\
    &Spheroid& 0 & 0\% & - & - & - & - \\
    &Intermediate Disk& 1 & 3\% & - & - & - & - \\
    &All& 31 & 100\% & - & - & - & - \\
    \hline
    \end{tabular}
    \caption{Statistical classification breakdown of the various categories for sources in our sample, based on the activity and morphological type. The total sample is classified based on NED classifications, while the isolated and interacting/merging samples are classified based on the WISE colours.$^{[**]}$ We could only classify 31 galaxy members of 15 out of 25 systems as better detailed in Section \ref{mergers-sect}. These 31 are reported in this table. Note some fractions may not add to 100\% due to round off error.}
    \label{classification_breakdown}
\end{table*}

\subsection{Luminosities and q-parameter estimates}\label{lum_cal_explained}

The ${1.28}\, \mathrm{GHz}$ total flux densities in  $\rm{W \,m^{-2}\, Hz^{-1}}$ were converted to $1.4 \, \mathrm{GHz}$ rest-frame spectral luminosities in $ \rm{W\,Hz^{-1}}$ using:
\begin{equation}
    L_\mathrm{1.4 \, GHz}= \frac{4 \pi D_\mathrm{L}^2}{(1+z)^{1+\alpha}} \left( \frac{1.4}{1.28} \right)  ^\alpha S_{1.28 \, \mathrm{GHz} } 
    \label{L14}
\end{equation}
where $z$ is the redshift, $D_\mathrm{L}$ is the luminosity distance in meters (m) and $\alpha = -0.7$ is the assumed spectral index (\citealt{Condon2018PASP..130g3001C}; Equation 56). Similarly, we converted all the VLA measurements taken at 1.425 or 1.49 GHz to $1.4 \, \mathrm{GHz}$ rest-frame spectral luminosities.

We used the redshift, the luminosity distance and the total Radio flux density for each source to make this conversion for the "Isolated Sample" (described in Section \ref{iso-results}). For the "Total Sample" (described in Section \ref{total_sample}), where the merging galaxies are considered as unresolved single IRAS sources, we assumed the redshift and the luminosity distance of one of the components and the \textit{Total Radio flux density} of the system to calculate the Radio luminosity of the system as a whole. Similarly, for those systems in the "Interacting/Merging Sample" (described in Section \ref{mergers-sect}) where the individual members can be separated as distinct radio sources, we use the \textit{Total Radio flux density} (i.e., sum of pixel values over an aperture covering all emission) for each component, as presented in Table 3 of \cite{Condon2021}, to calculate their radio spectral luminosity using Equation \ref{L14}. In cases where total flux densities were not available, we instead used the \textit{integrated radio flux densities} (i.e., flux estimated from a Gaussian model fitted to the component(s)), also provided in the same table. Note that for MCG -02-01-052, no direct total radio flux measurement is reported in \cite{Condon2021}. However, we estimate its flux to be roughly 5.2 mJy by subtracting the flux density of MCG -02-01-051 from the total flux density reported for the pair. The radio spectral luminosities and other properties of the galaxies in the Interacting/Merging Sample are reported in Table \ref{meerligs_props_mer}. 

In the study by \citet{Sanders2003}, the objects' recession velocities ($cz$) and comoving distances ($D_\mathrm{c}$) were reported based on a flat cosmology with $ H_\mathrm{0} = 75 \, \mathrm{km \, s^{-1} \, Mpc^{-1}}$, $\Omega_\mathrm{m}\, = \, 0.3$ and $\Omega_\mathrm{{\Lambda}}\, = \, 0.7$. We determined, corrected and transformed all the \cite{Sanders2003} measurements to match our assumed cosmology defined in Section \ref{intro} by using the following equation:
\begin{equation}
   L_\mathrm{TIR} = \frac{D_\mathrm{L}^2}{D_\mathrm{L,S03}^2} \times L_\mathrm{TIR,S03}
  \label{LIR}
\end{equation}
Here $D_\mathrm{L}$ is the luminosity distance calculated using our cosmology and $z_\mathrm{cmb}$ (as described in Section \ref{Redshift}), $D_\mathrm{L,S03}$ and $ L_\mathrm{TIR,S03}$ represent the \citet{Sanders2003} luminosity distance and total infrared luminosity, respectively. This method was used to determine the TIR luminosity for the "total" and "isolated" samples. For the "interacting/merging" sample, where the TIR luminosity from IRAS is unavailable for each system's component, we estimated the corresponding $L_\mathrm{TIR}$ using WISE MIR data (specifically the W$\rm 3$ and W$\rm 4$ bands). These estimates were based on the scaling relations in \citet{cluver2017} and the recently updated calibration from \citet{Cluver2024arXiv241013483C}. \citet{cluver2017} found that there is a tight scaling correlation between the mid-infrared and the $L_\mathrm{TIR}$ of nearby galaxies, and hence is a convenient method to estimate the dust-obscured star formation rate; see also \citet{Yao2020,yao2022} who employed WISE measurements, specifically using the W3 band as a proxy for TIR, to estimate "$q_\mathrm{TIR}$" parameter, for a large sample of galaxies in the GAMA G$\rm 23$ field. 

Despite using different methods to estimate the infrared luminosities, we made sure that we could compare results between samples by comparing the $L_\mathrm{TIR}$ from IRAS and the $L_\mathrm{TIR}$ from WISE for those isolated objects for which we have both estimates. This comparison is reported in Appendix \ref{IRAS_WISE_lum_comp_appendix} and shows good consistency between the two measurements, with $71\%$ of sources agreeing within ±0.25 dex. Therefore, we expect to be able to compare our results for the interacting/merging sample with the rest of the samples without any issues arising from the different luminosity estimation methods employed.

Using Equations \ref{L14} and \ref{LIR}, we computed the infrared-radio correlation for the total and isolated samples. The infrared/radio correlation was originally parameterised in terms of flux density ratio in the IRAS far-infrared (FIR) bands, specifically at $\rm 60\, \micro m$ and $\rm 100\, \micro m$. This quantity has come to be widely known as $q_{\text{FIR}}$ \citep{Helou1985, yun2001}. However, here we utilise the correlation, parameterised in terms of total infrared/radio luminosities. This choice is mainly because the infrared luminosities we employ are taken from the work of \citet{Sanders2003}, which incorporates data from all four IRAS bands. This is the main advantage of RBGS, as often most galaxies are too faint for IRAS to detect at 12 and 25 $\mu$m. Hence, the correlation links to the total infrared luminosity (TIR) via the $q_\mathrm{TIR}$ parameter as
\begin{equation}\label{q_equation}
      q_\mathrm{TIR} = \rm{log} \left(  \frac{L_\mathrm{TIR}\, (W)}{3.75 \times 10^{12}\, \rm{Hz} }  \right) - log\left( \frac{L_{1.4 \,\rm{GHz} }}{\rm {W \,Hz^{-1}}} \right) 
\end{equation}
with $ L_\mathrm{TIR}$ (W) being the dust continuum luminosity integrated over the rest-frame wavelength region ($\rm 8 - 1000 \, \mu m$) in the infrared, calculated using

\begin{equation}
F_\mathrm{TIR} = 1.8 \times 10^{-14}(13.48\, \mathit{S}_{12} + 5.16\, \mathit{S}_{25} +2.58\, \mathit{S}_{60} + \, \mathit{S}_{100})
\end{equation}
in units of $\rm W\, m^{-2}$, where the quantities $\rm \mathit{S}_{12}$, $\rm \mathit{S}_{25}$, $\rm \mathit{S}_{60}$ and $\rm \mathit{S}_{100}$ represent the IRAS flux densities in Jy at $\rm 12,\, 25,\, 60,\,$ and $\rm 100\, \mu m$, respectively. Then, in Equation \ref{q_equation},  $L_\mathrm{TIR}$ is equal to $4 \pi D_{\mathrm{L}}^2$$ F_\mathrm{TIR}$, $ L_\mathrm{1.4 \, \rm{GHz}}$ is the radio spectral luminosity at $\rm 1.4$ GHz and $\rm 3.75 \times 10^{12} \,Hz$ is the central Infrared band frequency such that $q_\mathrm{TIR}$ is dimensionless. For the "interacting/merging" sample, we first estimated the total infrared (TIR) luminosities (\( L_\mathrm{TIR'} \)) using the WISE 12\,\(\mu\)m band via the relation defined in Equation~\ref{cluver_LTIR_proxy}, adopted from Equation~1 of \citet{cluver2017}. The radio spectral luminosities were then calculated from the total/integrated radio flux densities, converted to $1.4 \, \mathrm{GHz}$. Then, using these two quantities, we computed the \( q_\mathrm{TIR'} \) parameter. Note that the prime notation is used solely to distinguish these WISE-derived measurements from those obtained using IRAS data.

\begin{equation}
\rm{log}[ {\mathit{L}_\mathrm{TIR'} (L \mathrm{_\odot}) }] =
(0.889 \pm 0.018)\, \rm{log}[ \mathit{L}_{12\, \mathrm{\mu m}}(L \mathrm{_\odot})]+ (2.21 \pm 0.15)
\label{cluver_LTIR_proxy}
\end{equation}
Thus, for this subset of interacting/merging sources, $q_\mathrm{TIR'}$ is defined as 

\begin{equation}
    q_\mathrm{TIR'} = \rm{log} \left(  \frac{\mathit{L}_\mathrm{TIR'} (W) }{3.75 \times 10^{12}\, \rm{Hz} }  \right) - log\left( \frac{\mathit{L}_{1.4\, \rm{GHz} }}{\rm {W \,Hz^{-1}}} \right) 
\end{equation}

All the quantities described in this Section are reported in Table \ref{meerligs_props_all} for the "total" and "isolated" samples and in Table \ref{meerligs_props_mer} for the "interacting/merging" southern sample. The full tables in a machine-readable format are available at \url{https://doi.org/10.5281/zenodo.16889781}.

We note the nonlinear relation between $L_\mathrm{TIR'}$ and the MIR star formation indicator $L_\mathrm{12 \, \mu m}$ in equation \ref{cluver_LTIR_proxy}. If we define the purely infrared TIR/$\mathrm{12\, \mu m}$ flux ratio in a similar way to the FIR/radio flux-density ratio, we get:

\begin{equation}
    q* \equiv \mathrm{log} [L_\mathrm{TIR'}(L_\mathrm{\odot}) ]- \mathrm{log}[L_\mathrm{12 \, \mu m} (L_\mathrm{\odot}) ],
\end{equation}
then 
\begin{equation}
    \left< q* \right> = 2.21-(0.111 \pm 0.018) \mathrm{log}[L_\mathrm{12 \, \mu m}  (L_\mathrm{\odot})].
    \label{q_pure_mir}
\end{equation}
This will be explored in detail in Sections \ref{FIR/TIR_radi_corr_factor} and \ref{discussion}.

\subsection{Stellar masses and star formation rates}\label{phys_props_derived_wise}
We make use of WISE mid-infrared data to estimate the physical properties of our sources. The galaxies' global stellar masses ($M_\star$) are derived following the prescription outlined in \citet{jarrett2023}, assuming the \citet{Chabrier2003PASP..115..763C} initial mass function (IMF). Star formation rates (SFRs) are estimated using the method described in \citet{Cluver2024arXiv241013483C}, corrected to \citet{Chabrier2003PASP..115..763C} IMF. These SFRs ($\mathrm{SFR}_\mathrm{MIR,corr}$) are computed via an inverse-variance weighted combination of the SFRs derived from the $W3$ and $W4$ luminosities, with corrections applied for deficit in low-mass and dust populations. For full details, refer to \citet{Cluver2024arXiv241013483C}.

\section{Results}\label{results}

\subsection{Total Sample Properties}\label{total_sample}
The total sample includes all the isolated objects and the interacting/merging systems. Our southern sources span $0.00012 < z < 0.08825$, $ 7.53<\rm{log}(\mathit{L}_\mathrm{TIR}/L_\mathrm{\odot})<12.42$ and  $18.47<\rm{log}(\mathit{L}_\mathrm{1.4 \, GHz} \rm{W\,Hz^{-1}})<23.84$. The northern sources span  $0.0002<z<0.07413$, $8.09<\mathrm{log}(\mathit{L}_\mathrm{TIR}/L_\mathrm{\odot})<12.56$ and  $19.18<\mathrm{log}(\mathit{L}_\mathrm{1.4 \, GHz} \rm{W\,Hz^{-1}})<25.09$. In all the plots, isolated objects observed with MeerKAT (southern sample) are represented as circles, while those observed with the VLA (northern sample) are shown as squares. Interacting/merging objects are shown in red as triangles, pointing up for the southern sample and down for the northern sample.

\subsubsection{Observed Infrared/radio correlation}\label{firc-result-ned}
\begin{figure}
    \centering
    \includegraphics[width=0.44\textwidth]{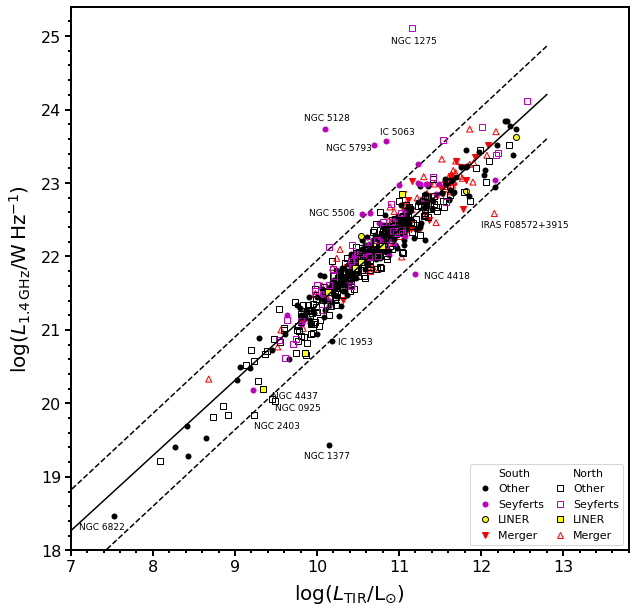}
    \caption{Radio spectral luminosity ($ L_{1.4 \, \rm{GHz}}$) versus the total infrared luminosity ($L_\mathrm{TIR}$) for our total sample. The sources are separated into different classes as described in Section \ref{nuclear_class}, with "Other" referring to sources that are neither Seyferts nor LINERs based on our classification. The red triangles indicate the interacting/merging systems. The black solid line represents the overall fit to all other objects in both samples and is defined by equation \ref{fit_linear_total}, while the black dashed lines indicate the $\rm \pm  3 \sigma$ deviations from the best fit.}
    \label{firc-q}
\end{figure}

In Figure \ref{firc-q}, we report the radio spectral luminosity against total infrared luminosity for both the southern and northern samples. The objects are separated into different classes based on literature, as explained in Section \ref{nuclear_class}. The plot shows a tight (a Pearson coefficient of 0.96) and nearly linear correlation spanning a wide range of luminosities. The best fit to this correlation has a slope that is close to unity (i.e., $1.02 \pm 0.01$). The best power-law fit to the observed infrared/radio correlation of our samples, with $L_\mathrm{TIR}$ scaled to $10^{10} \mathrm{L_{\odot}}$, is given by:

\begin{equation}\label{fit_linear_total}
\rm
    log \left(\frac{\mathit{L}_\mathrm{1.4 \, GHz}}{W \, Hz^{-1}} \right) = 1.02 \pm 0.01\times log\left(\frac{\mathit{L}_\mathrm{TIR}}{10^{10} L_\mathrm{\odot}}\right) + 21.34 \pm 0.01
\end{equation}

This fit was derived excluding the interacting/merging sources and those identified as Seyferts or LINERs with NED, to reduce the effects of AGN contamination and potential biases that could be introduced by mergers. We considered as outliers those objects that are $\rm 3 \sigma$ away from the rest of the sample. These objects are marked in the plot (identified with their names), and they were also not considered when deriving the fit. 

\begin{figure}
    \centering
    \includegraphics[width=0.44\textwidth]{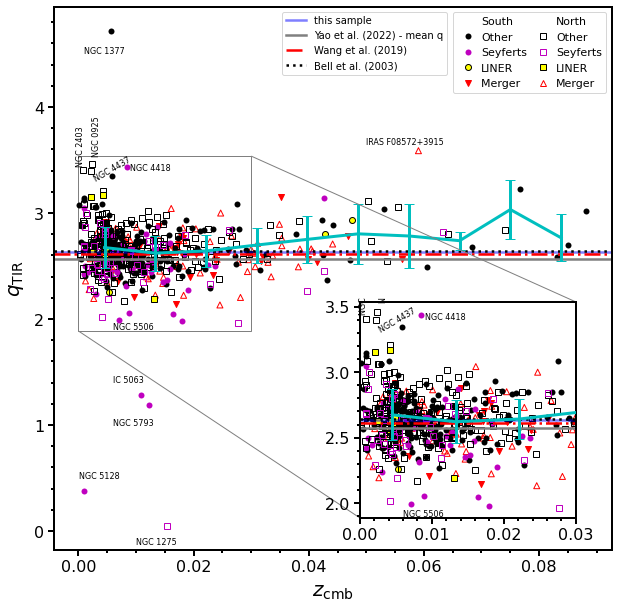}\\
    \hspace{0.01pt}
    \includegraphics[width=0.44\textwidth]{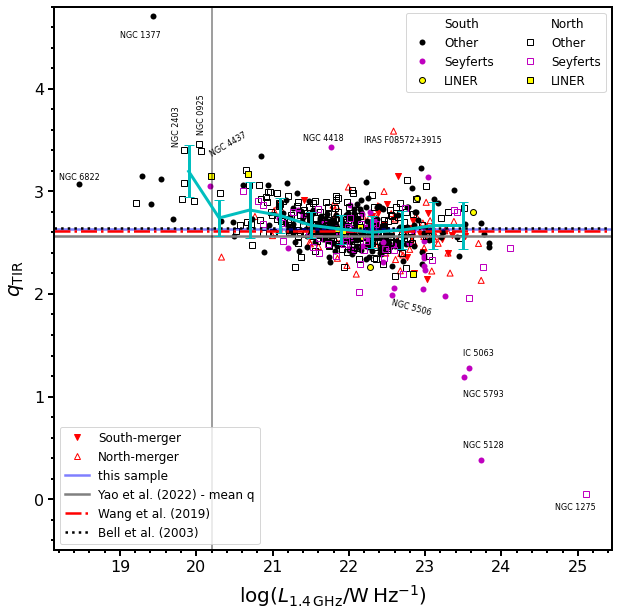}
    \caption{Distribution of $q_\mathrm{TIR}$ against redshift (top) and radio spectral luminosity (bottom) for our southern and northern samples. The sources are similarly separated into different classes as described in Figure \ref{firc-q}. In both panels, the blue line represents the median of the all-sky sample, excluding measurements of sources identified as LINERs/Seyferts. The other horizontal lines represent measurements from the literature that we use to compare with our sample, as in the legend. The vertical line depicts a region with fewer points, we only use objects with luminosities above this threshold (i.e., $\rm{log}(L_{1.4\, GHz}/\rm{W\, Hz^{-1}})\, >\, 20.2$) to compute the median of our sample. Finally, the cyan line illustrates the binned average for the redshift (top) and radio spectral luminosity (bottom). } 
    \label{qtir-z-lrad-total}
\end{figure}

Examining the "$q$-parameter" makes it easier to spot any non-linearity between the infrared and radio correlation by more clearly highlighting the existing dispersion. In Figure \ref{qtir-z-lrad-total}, we show the distribution of $q_\mathrm{TIR}$ derived for this sample against redshift ($z$; top panel) and radio spectral luminosity ($L_{1.4 \,\rm{GHz}}$; bottom panel). The median $q_{\mathrm{TIR}}$ values for the non-Seyfert/LINERs sources in the southern and northern hemisphere objects are $2.62 \pm 0.01$ with a scatter of 0.17 and $2.65 \pm 0.02$ with a scatter of 0.18, respectively. The quoted uncertainties on the medians were derived via bootstrap resampling and are consistent with the expected statistical error of $\sim 0.17/\sqrt{N}$, confirming the robustness of the estimates. Although a Student’s t-test indicates a marginally significant difference between the two samples ($t = -1.99$, $p = 0.04776$), this falls within the range of expected statistical variation and likely does not impact the reliability of the median values. The overall $q_{\mathrm{TIR}}$ median value for the whole sample is $2.63 \pm 0.01$, with a scatter of 0.17. 
 
We compare this total $q$ median with findings from previous studies that examined the $q_\mathrm{TIR}$ in the local universe and one where the radio flux density used to compute $q_\mathrm{TIR}$ was measured in the same frequency as in our study, but for sources at a slightly higher redshift. \cite{bell2003} investigated the infrared/radio correlation using a diverse sample of local galaxies and found a median $q_\mathrm{TIR}$ value of $\rm 2.64 \pm 0.02$ (overplotted on Figure \ref{qtir-z-lrad-total} as a black dotted line) with a scatter of 0.26. \cite{wang2019}, studied this correlation using IRAS and LOFAR at $\rm 150$ MHz and found a median $q_\mathrm{TIR}$ value of $\rm 2.61$ (represented with red line) with a scatter of 0.30. Lastly, \cite{yao2022} found a median of $2.57 \pm 0.23N^{-0.5}$ (where N is the number of galaxies in the sample) using MeerKAT and WISE-derived total infrared luminosity. We find that our results are comparable with \cite{bell2003, wang2019, yao2022}.

Most of the objects largely overlap in the same $q_\mathrm{TIR} \,-\, z$ parameter space (Figure \ref{qtir-z-lrad-total} upper panel). The Seyfert population tends to deviate more compared to the other classes. This is expected, especially for radio-loud sources (e.g., NGC 1275 and NGC 5128), where the excess radio emission from the AGN significantly contributes to the overall emission, leading to deviations from the infrared-radio correlation. We also show the binned average of the data (overplotted in cyan), and found that at lower $z$, the line slightly decreases and quickly reaches a plateau, with the error bars being consistent and uniform. However, due to the limited redshift range of our sample, we cannot investigate this further, as it restricts our ability to detect any potential evolution with redshifts, if present. Therefore, we do not discuss this further.

We proceed to show $q_\mathrm{TIR}$ as a function of $L_{1.4\, \mathrm{GHz}}$ (Figure \ref{qtir-z-lrad-total} lower panel). The different classes of objects are still found to be largely overlapping with one another. At lower radio spectral luminosities, the objects tend to have higher $q_\mathrm{TIR}$ values and appear more spread out. This dispersion might be accentuated by the limited number of data points. Hence, we applied a cutoff at $ \mathrm{log} [L_\mathrm{1.4\, GHz} \mathrm{(W \, Hz^{-1})} ] \,=\, 20.2$, indicated as a grey vertical line on the plot, below which we have fewer (i.e., $\rm 13 $) sources. We use only the objects with luminosities greater than this threshold to compute our median $q_\mathrm{TIR}$. Within $21 < \mathrm{log}[L_\mathrm{1.4 \, GHz} \rm{(W\,Hz^{-1})}] < 23.6$ the objects are mostly clustered around the median value. We observe a slight decline going from lower to higher luminosities before the trend reaches a plateau. This trend can be clearly observed in the binned average line, shown as a cyan line. \cite{Matthews2021} measured $q_\mathrm{FIR}$ as a function of $\rm{log}(L_\mathrm{1.4\,GHz})$ from a large volume-limited sample of $4.3 \times 10^3$ 2MASX nearby SFGs. They noted a similar trend (i.e., decline followed by flattening; see their Figure 10) to what we observe between luminosity and the $q_\mathrm{TIR}$ parameter, with their decline occurring below $\mathrm{log}(L_{1.4\,\rm{GHz}}) =  22.5$, albeit offset by $ q_{FIR} - q_\mathrm{TIR} \approx - 0.36 $ (as described in Subsection \ref{FIR/TIR_radi_corr_factor}), compared to our findings. 

At higher radio spectral luminosities ($ \mathrm{log} [L_\mathrm{1.4\,GHz} \rm{(W\,Hz^{-1})}] > 23)$ the objects classified as AGNs exhibit lower $q_\mathrm{TIR}$ values. A total of $\rm 12$ objects (5 of which are AGNs) in our sample deviated more (> 3$\sigma$) from the correlation. These outliers are discussed in detail in Section \ref{outies}.

\subsection{Isolated Sample}\label{iso-results}
We further analysed the correlation using the classifications based on the mid-infrared colours (as described in Section \ref{nuclear_class}) for the isolated objects. This subset contains $\rm 273$ and $\rm 249$ objects in the southern and northern hemisphere, respectively, yielding a total of $\rm 522$ sources. 

\subsubsection{WISE colours}\label{wise_col}
\begin{figure*}
    \centering
    \includegraphics[width=0.7\textwidth]{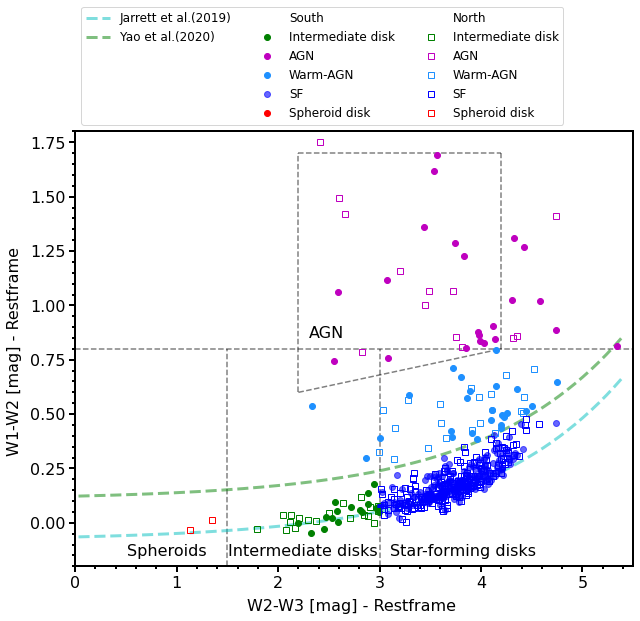}
    \caption{WISE mid-infrared k$-$corrected colours: $W1-W2$  versus $W2 - W3$ colour-colour diagram for the isolated sources in the southern (filled circles) and northern (empty squares) samples. The dashed grey lines and box delineate the location of the different classes of objects on the WISE colour-colour plane \citet{Jarrett2011,stern2012,jarrett2017}. The dashed cyan line indicates the star-formation fit defined by equation 1 in \citet{Jarrett2019}, and the dashed green line represents the $\rm 2 \sigma$ offset defining the lower threshold for "warm AGN" as defined by \citet{Yao2020}. The sources are further separated (by colour) from each other based on the location they occupy in this colour space.}
    \label{fig:wise-2col-iso}
\end{figure*}

Figure \ref{fig:wise-2col-iso} shows the WISE $k$-corrected colour-colour diagram, ($W1-W2$) against ($W2 - W3$), for our sample. The dashed box (taken from \citealt{Jarrett2011}) indicates the expected location for luminous AGNs, while the $W1-W2= 0.8 \, \mathrm{mag}$ line is a conservative limit for AGNs \citep{stern2012}; galaxies above this line are classified as WISE AGNs and are often in the Quasars (QSO) class. All the classes used to delineate the objects are represented in dashed lines on the graph, and the different populations are shown with colours or symbols. These include SF galaxies, AGNs, intermediate disks (i.e., the "green valley" galaxies), spheroids and Warm-AGNs (i.e., a subset of lower power AGNs relative to the host emission).  We also show the star formation sequence (cyan dashed line) computed by \cite{Jarrett2019} and a 2$\rm \sigma$ offset from the SF sequence  (green dashed line) adopted from \citet{Yao2020}. This offset delineates the region where the warm-AGN population resides \citep{Yao2020}. We refer the reader to \citet{jarrett2023} for further details about the corrections.

As expected, SF galaxies remain the dominant class in the sample, even with this classification scheme. Although we used only SF galaxies to compute the infrared/radio correlation for our isolated sample, we included all other galaxy classes in our plots for completeness.

\subsubsection{Infrared/radio correlation}\label{FIR/TIR_radi_corr_factor}
\begin{figure*}%
    \centering
    \includegraphics[height=0.49\textwidth, width =0.49\textwidth]{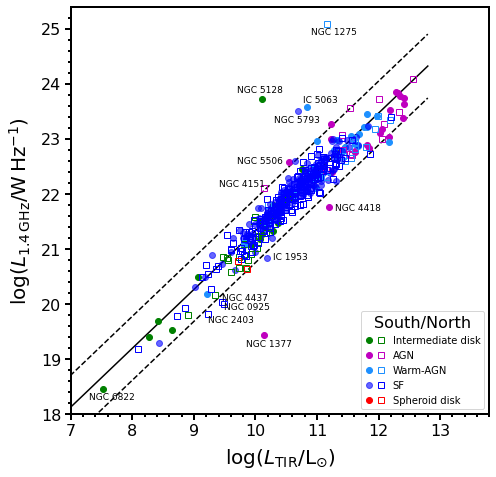}
    \hspace{0.06pt}
    \includegraphics[height=0.49\textwidth, width =0.49\textwidth]{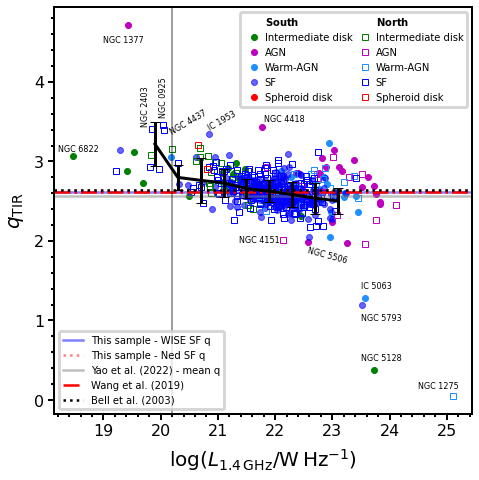}
    \caption{Total infrared luminosity ($L_\mathrm{TIR}$) against the radio spectral luminosity (left), and the ratio $q_\mathrm{TIR}$ versus radio spectral luminosity (right) for the isolated sources only in our southern and northern samples. The sources are separated into different classes based on the WISE classification scheme described in Figure \ref{fig:wise-2col-iso}. In the left panel, the black line is the fit to the objects classified as star-forming (SF) in both samples and is defined by the equation \ref{fit_linear-iso}. In the right panel, the vertical line marks the threshold used to delineate regions with a limited number of data points. Specifically, only the sources to the right of this line are considered when deriving the statistical information of the sample. Finally, the thick black solid line represents the binned average of the data, with the error bars indicating the scatter.}
    \label{firc-q-iso}%
\end{figure*}

Figure \ref{firc-q-iso} shows the radio spectral luminosity against the total infrared luminosity (left panel) and the ratio $q_\mathrm{TIR}$ versus radio spectral luminosity (right panel) for the isolated objects in this sample. In the left panel, the black line represents a fit to the SF galaxies in both the southern and northern samples of isolated galaxies and is defined by equation \ref{fit_linear-iso}, but also scaled to $10^{10} \mathrm{L_{\odot}}$. Even here, we still observe a tight correlation (with a Pearson coefficient of 0.96) among the SF galaxies. 
\begin{equation}
    \mathrm{log} \left (\frac{L_{1.4\,\rm{GHz}}}{\rm{W \, Hz^{-1}}} \right) = 1.04 \pm 0.02 \times \mathrm{log} \left(\frac{L_\mathrm{TIR}}{10^{10}\mathrm{L}_\mathrm{\odot}}\right) + 21.37\pm 0.01
    \label{fit_linear-iso}
\end{equation}

This is also illustrated in the right panel of Figure \ref{firc-q-iso}, which shows the ratio $q_\mathrm{TIR}$ as a function of the radio spectral luminosity. Most of the sources are clustered around the same median $q_\mathrm{TIR}$ (normalised median absolute deviation (NMAD) = 0.14). The median $q_\mathrm{TIR}$ value for the SFGs in the southern and the northern samples based on this WISE classification scheme is $\rm 2.60 \pm 0.01 $ with scatter of 0.15 and $\rm 2.62 \pm 0.01$ with scatter of 0.17, respectively, bringing the overall median for the all-sky to $\rm 2.61 \pm 0.01$ with scatter equal to 0.16. We further compare this result with findings from literature (\citealt{bell2003, wang2019, yao2022}) and still find that the $q_\mathrm{TIR}$ values are consistent with each other. All these are represented as horizontal lines in the right panel.

Similarly to Subsection \ref{firc-result-ned}, to check for any underlying trends in this subset, we compute the binned average, which is presented in the right panel as a black line. Above  $\rm{log} [\mathit{L}_{1.4\,\rm{GHz}} \rm{(W\,Hz^{-1}})] \sim 20.2\,$ (in Figure \ref{firc-q-iso}), we observe that the binned $q_\mathrm{TIR}$ trend initially remains constant before beginning a steady decline at $\rm{log}[L_{1.4 \, \rm{GHz}} \rm{(W\,Hz^{-1}})] > 21$, as we move to higher radio spectral luminosities. When considering the WISE SF sources with $\log[L_{1.4,\rm{GHz}} \rm{(W \, Hz^{-1})}] > 20.2$, the binned averages of $\log L_{1.4,\rm{GHz}}$ and $q_{\rm{TIR}}$ exhibit a moderate negative linear relationship, with a Pearson correlation coefficient of -0.38 (p $\approx 1.02 \times 10^{-14}$), indicating a statistical significance.

\begin{figure}
    \centering
    \includegraphics[width =0.49\textwidth]{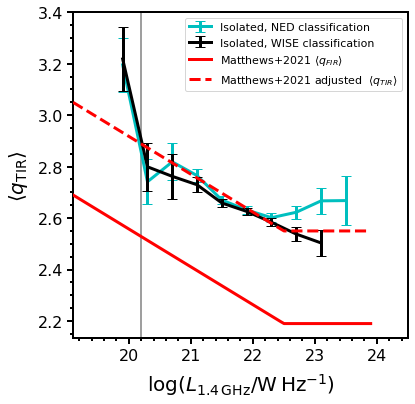}
    \caption{Comparison of the binned average of the radio spectral luminosity for the objects classified using both the NED and mid-infrared colours for the isolated objects. The error bars represent the standard error on the mean $q_\mathrm{TIR}$ per bin. The red line represents the best fit defined by equation \ref{matthews_bestfit} from \citet{Matthews2021}, while the dashed red line represents its adjustment from $q_\mathrm{FIR}$ to $q_\mathrm{TIR}$ for determining the correction factor between TIR and FIR measurements. The vertical line is as explained in Figure \ref{firc-q-iso}.} 
    \label{rolling_mean_com}
\end{figure}

We compare median outcomes derived here to those obtained using the NED classifications for the total sample in Figure \ref{rolling_mean_com}. On average, at intermediate radio spectral luminosities, both classification schemes result in the decline of $\left<q_\mathrm{TIR} \right>$ with increasing radio spectral luminosities for the SF populations. However, at higher luminosities ($\rm{log} [\mathit{L}_{1.4\,\rm{GHz}} \rm{(W\,Hz^{-1}})] \approx 22.3\,$), the NED-classified sample exhibits a slight increase in $\left<q_\mathrm{TIR} \right>$, which plateaus in the highest luminosity bin. This trend is likely driven by AGN contamination that is not effectively excluded.  In contrast, the WISE-classified sample maintains a declining trend, indicating that WISE-based diagnostics more effectively separate AGN from star-forming galaxies at these luminosities. The implications of this declining trend are discussed in Section \ref{discussion}. This declining feature is consistent with the trend observed by \citet{Matthews2021} defined by equation \ref{matthews_bestfit} (shown as the red solid line in Figure \ref{rolling_mean_com}), although it is offset by the difference between FIR and TIR. 
\begin{equation*}
 q_\mathrm{FIR} = 2.19 - 0.147 [\mathrm{log} (L_\mathrm{1.4 \, GHz}) - 22.5] \, \mathrm{when} \, \mathrm{log}( L_\mathrm{1.4 \, GHz}) < 22.5 
\end{equation*}
\noindent and
\begin{equation}
q_\mathrm{FIR}=2.19  \, \mathrm{when} \, \mathrm{log}( L_\mathrm{1.4 \, GHz}) > 22.5\\
\label{matthews_bestfit}
\end{equation} 
Interestingly, upon transforming the \citet{Matthews2021}  $\left < q_\mathrm{FIR} \right>$  to $\left < q_\mathrm{TIR} \right>$, we find that it aligns with our findings as depicted by the red dashed line in Figure \ref{rolling_mean_com}. 

To quantify the difference between our $\left < q_\mathrm{TIR} \right>$ and Matthews' $\left < q_\mathrm{FIR} \right>$, we carried out a formal fit analysis for their "best fit" to our data, assuming that $\left <q_\mathrm{TIR} \right> - \left <q_\mathrm{FIR} \right>$ is nearly independent of $\mathrm{log}( L_\mathrm{1.4 \, GHz})$; which implies that  $\left < q_\mathrm{TIR} \right>$ varies with radio luminosity as $-0.147\, (\mathrm{log} L_\mathrm{1.4 \, GHz})$ just as $\left < q_\mathrm{FIR} \right>$ does. We find that the difference between their $\left < q_\mathrm{FIR} \right>$, and our $\left < q_\mathrm{TIR} \right>$ is $\mathrm{dex}(0.36 \pm 0.04)$, implying that  $\left < L_\mathrm{TIR}/ L_\mathrm{FIR}\right> \approx 2.29$, slightly higher than the factor reported in \citet{bell2003}. \citet{bell2003} compared $q$-values between TIR and FIR luminosities and found $\left <L_{TIR}/L_{FIR} \right> = \mathrm{dex}(2.64 - 2.36)= \mathrm{dex}(0.28) \approx 1.91$. However, it is important to note that his findings did not agree with those from \citet{Sanders1996}, and his sample selection, which required all galaxies to have published FUV flux densities, introduces bias, rendering the dex($0.28$) result somewhat uncertain. Therefore, our estimate of dex($0.36 \pm 0.04$) should be the reference of choice for quantifying the difference between TIR and FIR luminosities because (1) we used a clean subsample of SF galaxies using WISE colours and (2) the RBGS galaxy sample is very bright, so the radio and infrared flux-density errors are small, rendering it more reliable. The adjusted \citet{Matthews2021} best fit in terms of the total infrared emission is defined by equation \ref{total_bestfit} below, shown as the red dashed line in Figure \ref{rolling_mean_com}.

\begin{equation*}
 q_\mathrm{TIR} = 2.55 - 0.147 [\mathrm{log} (L_\mathrm{1.4 \, GHz}) - 22.5] \, \mathrm{when} \, \mathrm{log}( L_\mathrm{1.4 \, GHz}) < 22.5 
\end{equation*}
\noindent and
\begin{equation}
q_\mathrm{TIR}=2.55  \, \mathrm{when} \, \mathrm{log}( L_\mathrm{1.4 \, GHz}) > 22.5\\
\label{total_bestfit}
\end{equation}

Notably, the $-0.147$ slope in this equation is close to the logarithmic slope $-0.111$ of the TIR/MIR $\left<q*\right>$ from Equation \ref{q_pure_mir} in Section \ref{lum_cal_explained} and it is significant at $0.111/0.018 \approx 6 \sigma \,$ level. The fact that both the pure TIR/MIR ratio and the TIR/Radio ratio inferred from the FIR/Radio ratio by \cite{Matthews2021} are nonlinear by the same amount is significant and may help us understand the nonlinearity in both these correlations. This is discussed in detail in Section \ref{discussion}.

\subsection{Interacting/Merging Sample}\label{mergers-sect}
Thanks to the sensitivity and resolution of instruments such as MeerKAT and WISE, we are now able to probe the individual properties of these interacting/merging objects to gain more insight into their properties. It is worth noting that we excluded the northern sample from further analysis of the interacting/merging systems, as their VLA observations do not typically resolve the multiple radio sources within each system. However, they are still accounted for in the total sample. Similarly, 10 systems out of 25 in the southern sample are already at an advanced merger stage and were therefore also excluded from detailed analysis. This left 15 systems from the southern sample that are still in the interaction phase. Accordingly, we further analysed the properties of these systems, as our MeerKAT observations effectively resolved their individual components. In total, this constitutes 31 distinct objects, consisting of 14 galaxy pairs and one triple system.

\subsubsection{WISE information}\label{mer-wise-cols}
\begin{figure*}
    \centering
    \includegraphics[width=0.7\textwidth]{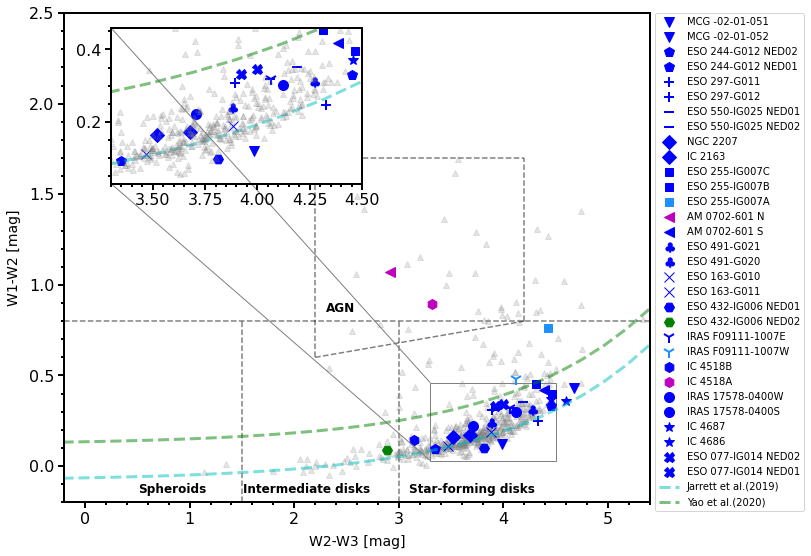}
    \caption{WISE mid-infrared k-corrected ($W1-W2$) colour versus ($W2 - W3$) colour diagram for the interacting/merging sources in the southern sample. All the delineations are as described in Figure \ref{fig:wise-2col-iso}, except that the sources with the same sample symbol represent objects interacting/merging. The grey triangles in the background represent the isolated sample. We note that galaxies in the same system often exhibit similar mid-infrared colours; see Section~\ref{mer-wise-cols} for more details.} 
    \label{wise_colours_mer}
\end{figure*}

Similarly to Section \ref{iso-results}, this sample was also split into different classes using the WISE colour-colour diagram (see Figure \ref{wise_colours_mer}). In addition, objects belonging to the same system are represented with the same symbol. For instance, interacting/merging sources IC 4518A/B (i.e., IC 4518A and IC 4518B) are both denoted by hexagons, with colour coding indicating their positions in the colour-colour space. It is interesting to see that some of the interacting objects/components lie in different regions of the WISE colour-colour space. For example, the galaxy pair IC $4518$ has one member of the pair with redder mid-infrared colours, indicating that it is dominated by star formation activity. In contrast, the other member is primarily dominated by emission from an AGN. The images of the two systems displaying this kind of behaviour are provided in Appendix \ref{diff_mech}. 

We note that most of the interacting/merging systems appear to exhibit mid-infrared colours more similar to each other than expected by random chance. To investigate this further, we compared their mid-infrared colours with those of isolated systems. Despite the apparent overlap between interacting/merging and isolated galaxies in WISE colour–colour space, a Kolmogorov–Smirnov test reveals a statistically significant (statistic = 0.446, p-value < 0.001) difference in their $W2 - W3$ colour distributions for sources of similar luminosities. This is consistent with expectations (i.e., for gas-rich sources), as galactic interactions/mergers can supply fuel for, and even trigger, star formation, leading to their bluer MIR colours (e.g., \citealt{Darg2010MNRAS.401.1552D, Alonso2010arXiv1001.4605A, Alonso2012A&A...539A..46A}). This aligns with \citet{Larson1978ApJ...219...46L}, who noted that peculiar galaxies tend to extend toward bluer colours due to bursts of star formation. Additionally, our findings corroborate those of \citet{Patton1997ApJ...475...29P}, who observed that galaxies undergoing interactions or mergers exhibit very blue rest-frame colours.

As these interacting/merging galaxy pairs were not resolved by IRAS, we employed their WISE mid-infrared properties to derive their total infrared luminosities, as explained in Section \ref{lum_cal_explained}. Although they were resolved by MeerKAT, estimating individual flux densities of the component galaxies was difficult in advanced, overlapping mergers. Hence, we were only able to perform this analysis for $\rm 28$ out of the $\rm 31$ resolved components. In total, 3 sources are merging, and the remaining 28 sources are still interacting. Of these 28, 23 are dominated by star formation (SF). Among these 23 sources, 9 systems have all of their resolved components dominated by star formation and the remaining 6 systems are dominated by different mechanisms. The full statistical classification breakdown for this sample is presented in Table \ref{classification_breakdown} and all the general properties of these sources are reported in Table \ref{meerligs_props_mer}. 

\begin{table*}
 \caption{Properties of the IRAS RBGS Southern Interacting/Merging Sample. The complete table is available on Zenodo at \url{https://doi.org/10.5281/zenodo.16889781}.}
 \label{meerligs_props_mer}
 \begin{tabular}{lllllllll}
  \toprule
  \hline
  Name & RA ($J2000$) & DEC ($J2000$)  & $z_\mathrm{helio}$ & $ z_\mathrm{cmb}$  & $D_\mathrm{L}$ & $ \rm{log}(L_\mathrm{TIR'})$ & $ L_\mathrm{1.4 \, GHz}$  & $q_\mathrm{TIR'}$ \\
    & $\rm hh:mm:ss$ & $\degr:\farcm:\farcs$ &  &  & $\mathrm{Mpc}$ & $\mathrm{L_{\odot}}$ & $\mathrm{W\,Hz^{-1}}$ & \\
    (1) &  (2) &  (3)  & (4) & (5) & (6) & (7) & (8)& (9)  \\
  
  \midrule
MCG -02-01-052 & 00 18 49.99 & -10 21 29.9 & 0.02733(0) & 0.0262 & 114.62 & 10.50 & 7.62e+21 & 2.63 \\ 
MCG -02-01-051 & 00 18 50.88 & -10 22 36.6 & 0.0271(0) & 0.02597 & 113.59 & 11.58 & 5.76e+22 & 2.82 \\ 
ESO 244-G012 NED01 & 01 18 08.26 & -44 27 58.9 & 0.02114(0) & 0.0205 & 89.31 & 10.25 & 6.50e+21 & 2.45 \\ 
ESO 244-G012 NED02 & 01 18 08.41 & -44 27 42.2 & 0.02093(0) & 0.02029 & 88.38 & 11.54 & 4.72e+22 & 2.88 \\ 
ESO 297-G011 & 01 36 23.49 & -37 19 17.4 & 0.0173(0) & 0.01661 & 72.15 & 10.83 & 2.11e+22 & 2.51 \\ 
ESO 297-G012 & 01 36 24.24 & -37 20 25.3 & 0.01732(0) & 0.01662 & 72.19 & 10.76 & 7.40e+21 & 2.9 \\ 
ESO 550-IG025 NED01 & 04 21 19.99 & -18 48 39.9 & 0.0322(0) & 0.03199 & 140.55 & 11.11 & 5.28e+22 & 2.4 \\ 
ESO 550-IG025 NED02 & 04 21 20.04 & -18 48 57.7 & 0.03199(0) & 0.03178 & 139.60 & 10.84 & 3.32e+22 & 2.33 \\ 
NGC 2207 & 06 16 22.02 & -21 22 22.3 & 0.00922(0) & 0.00959 & 41.44 & 10.67 &  &  \\ 
IC 2163 & 06 16 27.30 & -21 22 28.0 & 0.00914(0) & 0.00951 & 41.09 & 10.87 &  &  \\ 
ESO 255-IG007A & 06 27 21.63 & -47 10 36.0 & 0.0394(0) & 0.03978 & 175.77 & 11.92 & 1.58e+23 & 2.74 \\ 
ESO 255-IG007B & 06 27 22.48 & -47 10 46.3 & 0.03868(0) & 0.03906 & 172.50 & 11.44 &  &  \\ 
ESO 255-IG007C & 06 27 23.06 & -47 11 02.4 & 0.03948(0) & 0.03986 & 176.13 & 10.78 & 3.58e+22 & 2.24 \\ 
AM 0702-601 N & 07 03 24.08 & -60 15 22.4 & 0.0313(0) & 0.03171 & 139.29 & 11.61 & 7.49e+22 & 2.74 \\ 
AM 0702-601 S & 07 03 28.47 & -60 16 44.3 & 0.0311(0) & 0.03151 & 138.39 & 11.51 & 4.78e+22 & 2.84 \\ 
ESO 491-G020 & 07 09 46.87 & -27 34 08.1 & 0.01008(0) & 0.01068 & 46.18 & 11.01 & 1.28e+22 & 2.91 \\ 
ESO 491-G021 & 07 09 49.89 & -27 34 30.0 & 0.00989(0) & 0.01049 & 45.35 & 10.26 & 8.64e+21 & 2.34 \\ 
ESO 163-G010 & 07 37 53.21 & -55 10 58.2 & 0.00933(0) & 0.00987 & 42.66 & 9.74 & 1.28e+21 & 2.64 \\ 
ESO 163-G011 & 07 38 05.38 & -55 11 27.4 & 0.01251(2) & 0.01251 & 54.17 & 10.74 & 1.68e+22 & 2.53 \\ 
ESO 432-IG006 NED01 & 08 44 27.21 & -31 41 51.4 & 0.01616(0) & 0.01707 & 74.17 & 10.70 & 1.45e+22 & 2.54 \\ 
ESO 432-IG006 NED02 & 08 44 28.93 & -31 41 31.1 & 0.01615(0) & 0.01706 & 74.13 & 10.54 & 8.05e+21 & 2.64 \\ 
IRAS F09111-1007W & 09 13 36.47 & -10 19 29.7 & 0.05479(0) & 0.05587 & 249.72 & 11.75 & 1.94e+23 & 2.47 \\ 
IRAS F09111-1007E & 09 13 38.86 & -10 19 19.6 & 0.05504(0) & 0.05612 & 250.88 & 11.52 & 1.30e+23 & 2.41 \\ 
IC 4518A & 14 57 41.24 & -43 07 55.5 & 0.01626(0) & 0.01686 & 73.25 & 11.06 & 8.94e+22 & 2.12 \\ 
IC 4518B & 14 57 44.95 & -43 07 55.0 & 0.01551(0) & 0.01611 & 69.95 & 10.51 & 1.04e+22 & 2.5 \\ 
IRAS 17578-0400W & 18 00 24.52 & -04 01 00.9 & 0.01424(0) & 0.01399 & 60.65 & 10.13 & 1.19e+22 & 2.06 \\ 
IRAS 17578-0400S & 18 00 34.01 & -04 01 43.3 & 0.01425(0) & 0.014 & 60.69 & 10.16 & 6.27e+21 & 2.37 \\ 
IC 4686 & 18 13 38.63 & -57 43 56.2 & 0.0165(0) & 0.01645 & 71.45 & 10.85 & 6.57e+21 & 3.05 \\ 
IC 4687 & 18 13 39.57 & -57 43 30.5 & 0.01726(0) & 0.01721 & 74.79 & 11.52 & 6.24e+22 & 2.74 \\ 
ESO 077-IG014 NED01 & 23 21 03.69 & -69 13 02.0 & 0.03803(0) & 0.03774 & 166.51 & 10.66 & 3.57e+22 & 2.12 \\ 
ESO 077-IG014 NED02 & 23 21 05.42 & -69 12 48.3 & 0.04174(0) & 0.04145 & 183.37 & 11.44 & 1.18e+23 & 2.37 \\ 
\bottomrule
 \end{tabular}\\
\begin{justify}
     NOTE:- Column (1): The Common names of the galaxy systems. An asterisk (*) indicate that the reported radio spectral luminosity was calculated using flux values we inferred (see Section \ref{lum_cal_explained} for detailed explanation), (2)-(3): The MeerKAT positions of the sources in J2000, taken from \citet{Condon2021} (4) to (6), same as Table \ref{meerligs_props_all} (7): The total infrared luminosity of each component estimated from WISE mid-infrared measurements. (8): The radio spectral luminosity calculated from integrated flux densities of the objects converted to 1.4 GHz from the MeerKAT $\rm 1.28$ GHz. Finally (9): is the corresponding infrared/radio correlation ($q_\mathrm{TIR'}$), with most sources having a formal error of less than $0.05$, except for a few that form a tail above this threshold. \\
\end{justify}
\end{table*}

\subsubsection{Infrared/radio correlation}
\begin{figure*}
    \centering
    \includegraphics[width=0.9\textwidth]{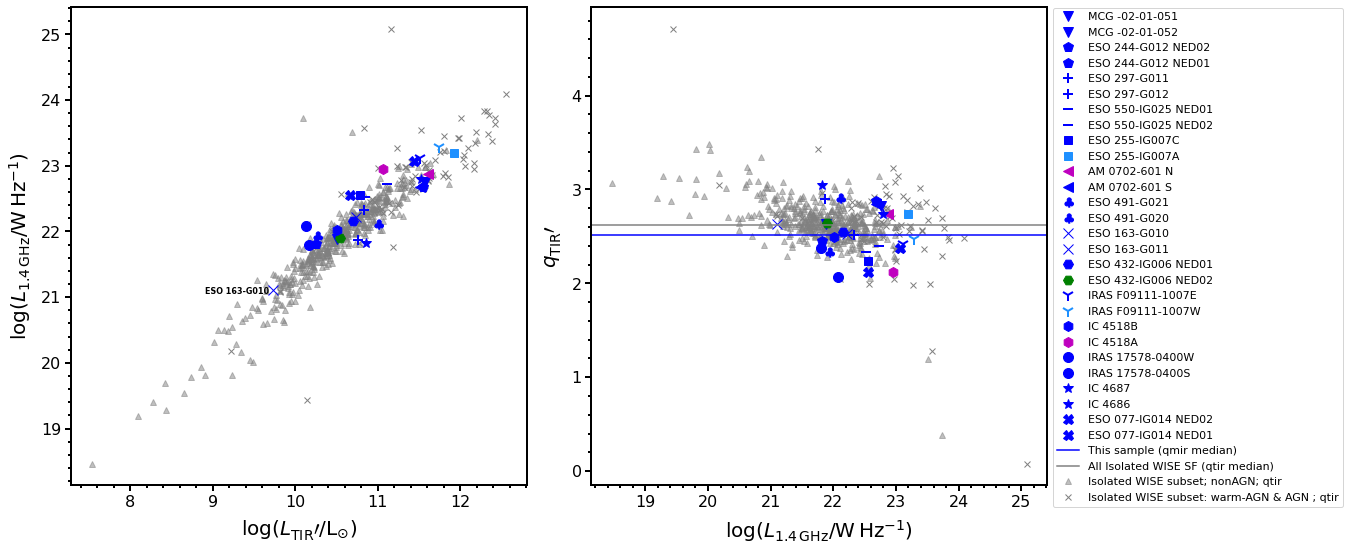}
    \caption{Total infrared luminosity ($L_\mathrm{TIR'}$) against the radio spectral luminosity (left), and the ratio $q_\mathrm{TIR'}$ versus the radio spectral luminosity (right) for the interacting/merging sources in the southern sample. The sources are separated into classes as in Figure \ref{wise_colours_mer}. In the right panel, the blue line represents the median of these interacting/merging objects classified as star-forming (SF). For comparison, the isolated sample, classified using the same scheme, is overplotted in grey (triangles represent WISE non-AGN sources, and crosses represent WISE AGNs, including warm-AGN class). The median for the star-forming galaxies is shown as a grey line. Among the galaxies plotted, ESO 163-G010 is faint in both infrared and radio spectral luminosity. We have highlighted this object on the left panel and discussed it in detail in Section \ref{outies}.
    }
    \label{firc-q-mult}
\end{figure*}

Figure \ref{firc-q-mult} shows the distribution of the radio spectral luminosity against the total infrared luminosity (left) and the ratio $q_\mathrm{TIR'}$ versus radio spectral luminosity (right) for the complex (i.e., MeerKAT resolved interacting) sources. Focusing on the interacting/merging galaxies, in the left panel of Figure \ref{firc-q-mult}, we notice a slight positive correlation (a Pearson coefficient of $0.809$) between the parameters, which still represents the infrared/radio relation. However, here we note these objects are a bit scattered, and the relation between their infrared and radio spectral luminosity is not as tight as the isolated subset (grey data points). This can also be seen on the right panel; the population is still largely scattered, spanning various values of the ratio $q_\mathrm{TIR'}$. We find that the median $q_\mathrm{TIR'}$ for this subset is $\rm 2.51 \pm  0.08$, with a scatter of 0.26. This median is represented by the blue line on the plot.

For comparison, we also overplotted the isolated star-forming population and other classes as grey triangles, with the SF median ($\rm 2.62 \pm 0.01$) indicated as a grey line, and also marked the AGNs with an x-symbol. On average, we find that interacting/merging galaxies tend to have lower $q_\mathrm{TIR'}$ values compared to their isolated counterparts and also exhibit greater scatter (NMAD = 0.27). A KS test further reveals that the variation in the $q$-parameter between interacting/merging and isolated galaxies of similar luminosities is marginally significant (KS statistic = 0.316, p-value = 0.024).

Taken together, these results support the idea that galactic collisions can introduce considerable scatter in the infrared–radio correlation (e.g., \citealt{Wunderlich1988A&A...206...47W, Murphy2013ApJ...777...58M, Donevski2015MNRAS.453..638D}) and provide direct observational evidence of increased dispersion in such systems.

While we observe that the merging sample exhibits more scatter and lower $q$-values compared to the isolated sample, it is important to acknowledge the limitations of our analysis. Firstly, though the majority of these systems are dominated by star formation, as shown in Figure \ref{wise_colours_mer}, we also note that they lie above the star formation sequence track. This shift towards warmer infrared colours as the track curves suggests that these merging systems could likely be experiencing bursts of star formation (e.g., \citealt{cox2008MNRAS.384..386C, Luo2014ApJ...789L..16L, Garay-Solis2023ApJ...952..122G}) or have contributions from a buried AGN (e.g., \citealt{Treister2012ApJ...758L..39T, Goulding2018PASJ...70S..37G, li2023ApJ...944..168L}), especially considering their higher radio luminosities. This likely occurs due to dynamical imbalances introduced by the merger, causing the gas to lose angular momentum and funnel toward the centre, where it can fuel a starburst, an AGN, or both. Alternatively, these systems might be hybrids, where the AGN contribution is not easily detectable.  

Nevertheless, we observe that these populations tend to occupy similar regions to those of the isolated sources. This suggests that despite the large scatter in the interacting/merging sample, the properties represented by the $q$-parameter and correlation are still consistent within statistical uncertainties. The interpretation of the large scatter and lower $q$-values is discussed in Section \ref{discussion}. The $\Delta q_\mathrm{TIR} = 0.1$ difference between the isolated and interacting/merging systems suggests that AGN activity triggered by mergers is weaker than star formation activity, with the median radio excess being $\mathrm{dex}(0.1) - 1 \approx 0.26$ as strong as the SFR activity.

\subsection{Host galaxy properties versus infrared/radio correlation }\label{host_props_mer_v_iso}
To determine whether the observed variations in the $q_\mathrm{TIR}$ are intrinsic to the galaxies, we investigate their relation to key host properties (i.e., total stellar mass $M_{\star}$ and star formation rate). In particular, we examine the relation between host galaxy properties and the infrared/radio correlation of our sources. These physical properties were derived as explained in Section \ref{phys_props_derived_wise}. 

\begin{figure*}
    \centering
    \includegraphics[width=0.9\linewidth]{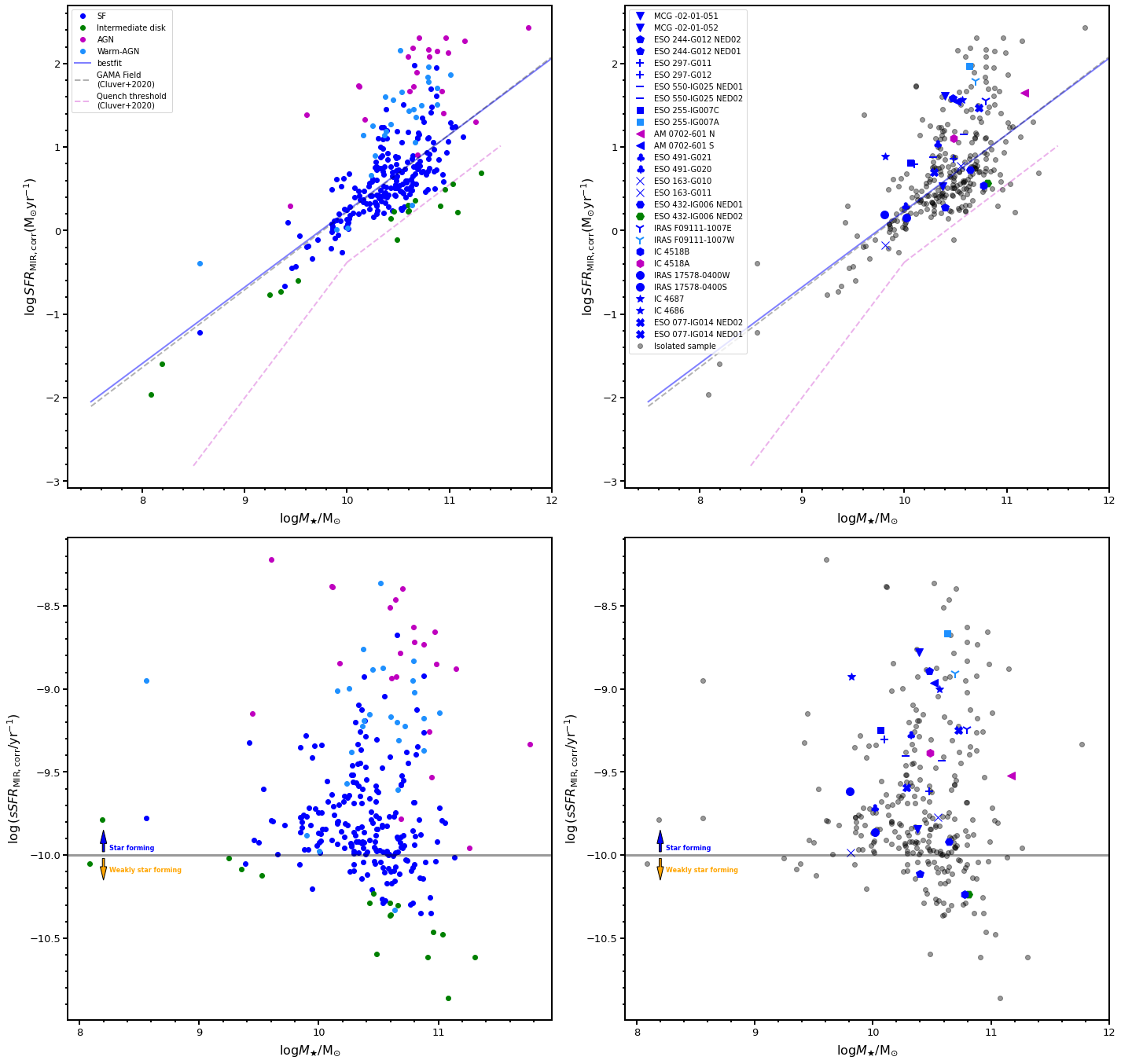}
    \caption{Top panel: WISE-derived star formation rates ($\mathrm{SFR}_\mathrm{MIR,corr}$) versus WISE stellar mass for the southern isolated (left) and interacting/merging (right) samples. The blue solid line represents the best fit to the isolated star-forming sources, while the dashed lines indicate measurements from the literature used for comparison. Bottom panel: Same as above, except now the WISE-derived specific star formation rate is on the ordinate. The grey lines mark the division between efficiently star-forming galaxies and those that are transitioning. The sources are separated into different classes as indicated in the legends. Overall, the SF populations in merging/interacting systems show greater scatter than isolated ones, reflecting the increased variability caused by mergers.} 
    \label{sfms-wise-derived-q}
\end{figure*}

In Figure \ref{sfms-wise-derived-q} top row, we plot the southern sample in the $SFR_\mathrm{MIR,corr}$ and $\mathrm{log} M_{\star}$ plane. These star formation rates are corrected for FUV emission and the SFR deficit in low-mass, low-dust galaxies. As expected in both panels, we observe a strong relation between the $SFR_\mathrm{MIR,corr}$ and $\mathrm{log} M_{\star}$ of our sources. Our best fit for the star formation main sequence (SFMS), is represented by the blue solid line on the plot and defined by Equation \ref{sfms-bfit} but scaled to $10^{10} \mathrm{M_\odot}$. This fit is consistent with the relation reported by \cite{Cluver2020ApJ...898...20C} within $1 \sigma$, after accounting for the scaling. The scatter around our SFMS is $\sim$0.28 dex, which falls within the typical range ($\sim$0.2–0.4 dex) reported in the literature for star forming galaxy (e.g., \citealt{Elbaz2007A&A...468...33E, Whitaker2012ApJ...754L..29W, Kurczynski2016ApJ...820L...1K, Thorne2021MNRAS.505..540T, Popesso2023MNRAS.519.1526P}). We note that the majority SF population follow the SFMS well. In the left panel, we also note that a few of the SF populations lie close to the quenching line of \cite{Cluver2020ApJ...898...20C}, which might be an indication that they are in the process of running out of fuel or transitioning. We also note that the majority of the intermediate disks also lie close to the quenching line and, in some cases, even below, suggesting they may also be transitioning. On average, the AGN and warm-AGN classes exhibit the largest scatter and seem to reside above SFMS, consistent with expectations.

\begin{equation}
    \mathrm{log} SFR_\mathrm{MIR,cor} (\mathrm{M_{\odot}} yr^{-1}) = (0.91 \pm 0.05) \mathrm{log} \left( \frac{M_{\star}}{10^{10}\mathrm{M}_\mathrm{\odot}}\right) + 0.24 \pm 0.03
    \label{sfms-bfit}
\end{equation}

In the right panel of Figure~\ref{sfms-wise-derived-q}, we show a similar distribution, this time for the interacting/merging sample. We now use this subset primarily for comparison with the isolated population, which is overplotted in the background as grey points. We acknowledge that our sample is biased toward galaxies with high SFRs (due to the infrared selection), and this may dilute any enhancement of SFR in merging systems relative to isolated systems. However, we note that the majority of the star-forming (SF) merger population follows the SFMS, although there is a slight tendency for them to lie above the main sequence. A KS test indicates statistically significant differences in their star formation rates ($SFR_\mathrm{MIR, corr}$; KS statistic = 0.400, p-value < 0.001) with respect to isolated galaxies of similar stellar masses.

In the bottom row of Figure~\ref{sfms-wise-derived-q}, we show the specific star formation rates ($SFR_\mathrm{MIR,corr}/\mathrm{log} \,M_{\star}$) versus stellar masses. We note that a small fraction of our sources exhibit lower activity (i.e., with $\mathrm{sSFR} < 10^{-10},\mathrm{yr}^{-1}$, indicated by the grey dashed line), while the majority remain actively star-forming and lie above this threshold. In both panels, we do not see any evidence of a particular trend (e.g., either systematic changes or correlations) between the sSFR and $M_\mathrm{\star}$ of our SF objects. On average, we note that the various populations seem to occupy different regions of this parameter space (left panel). The majority of the AGN and warm-AGN classes live in the high sSFR and stellar mass region, followed by the SF populations with intermediate sSFR and finally, the intermediate-disk covering the lower regions. However, we note that the higher sSFR values observed in these AGN hosts may be influenced by AGN contamination artificially boosting SFR estimates. As such, this represents a limitation of our analysis, and the derived SFRs for these AGN hosts should be interpreted with caution.

In the right panel, we note that the interacting/merging SF populations are scattered on this parameter space, and they occupy the same range as the isolated ones. Even though they still largely overlap with the isolated subset, they exhibit a large scatter on average. A KS test further confirmed that the differences in specific star formation rates between the isolated and interacting/merging populations of similar stellar masses are statistically significant (KS statistic=0.395, p-value=0.001). For the detailed statistical breakdown of the quantities of the SF sources in the isolated and interacting/merging samples, divided into three stellar mass bins, refer to Table \ref{binned_phys_props}.

\begin{table*}
\setlength{\extrarowheight}{5pt}
\centering
\caption{Comparison of the statistical properties (median $\pm$ NMAD) of the SF isolated and interacting/merging sources, binned according to stellar mass. }
\begin{tabular}{c|cccccc}
\hline
 & Stellar Mass Range & log$(M_\star/M_\odot)$ & log$(\mathrm{SFR}_{\mathrm{MIR,corr}})$ & log$(\mathrm{sSFR}_{\mathrm{MIR,corr}})$ & $W2-W3$ (mag) & $q_{\mathrm{TIR}}$ \\
\hline
\multirow{3}{*}{\textbf{Isolated}} 
& $9.81 - 10.13$  & $9.97 \pm 0.11$  & $0.15 \pm 0.18$  & $-9.77 \pm 0.14$  & $3.77 \pm 0.24$  & $2.62 \pm 0.17$ \\
& $10.13 - 10.45$ & $10.33 \pm 0.07$ & $0.50 \pm 0.26$  & $-9.81 \pm 0.24$  & $3.80 \pm 0.35$  & $2.63 \pm 0.14$ \\
& $10.45 - 10.77$ & $10.59 \pm 0.12$ & $0.69 \pm 0.23$  & $-9.95 \pm 0.19$  & $3.64 \pm 0.25$  & $2.58 \pm 0.12$ \\
\hline
\multirow{3}{*}{\textbf{Interacting/Merging}} 
& $9.81 - 10.13$  & $10.02 \pm 0.06$ & $0.79 \pm 0.14$  & $-9.30 \pm 0.56$  & $4.32 \pm 0.41$  & $2.34 \pm 0.40$ \\
& $10.13 - 10.45$ & $10.32 \pm 0.06$ & $0.88 \pm 0.51$  & $-9.40 \pm 0.65$  & $4.19 \pm 0.39$  & $2.54 \pm 0.36$ \\
& $10.45 - 10.77$ & $10.57 \pm 0.10$ & $1.01 \pm 0.57$  & $-9.53 \pm 0.57$  & $3.95 \pm 0.30$  & $2.53 \pm 0.19$ \\
\hline
\end{tabular} 
\label{binned_phys_props}
\end{table*}

Recent work by  \cite{Delvecchio2021} reported finding evidence of the dependence of the infrared/radio correlation on stellar mass, with more massive galaxies displaying a systematically lower $q$-values over a wide range of redshift $ 0.1 < z < 4$ and stellar mass ($8.5 < \mathrm{log}(M_{\star}/\mathrm{M_{\odot}}) < 11.5$). Since our sample comprises galaxies at the lower redshift end ($z<0.1$), we decided to check whether our sources will also display this behaviour to give a perspective of the nearby Universe.

\begin{figure*}
    \centering
    \includegraphics[width=\linewidth]{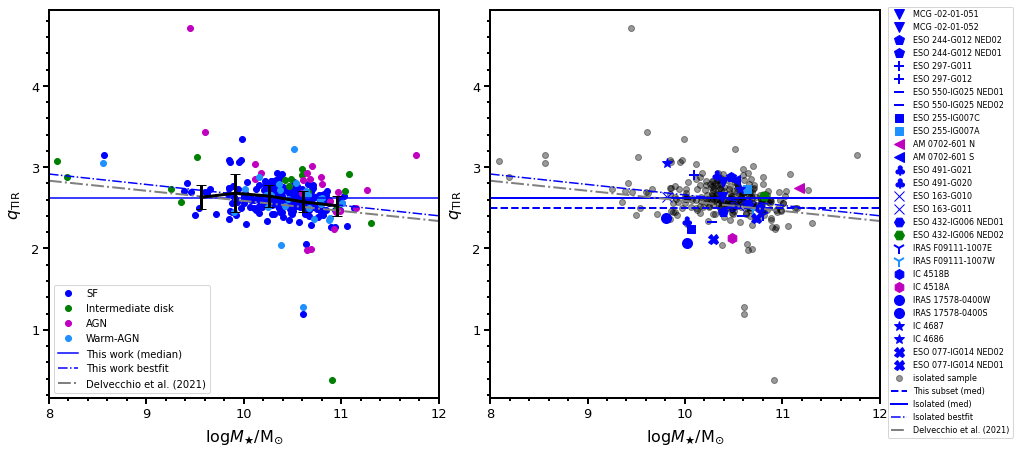}
    \caption{Infrared/radio correlation against the stellar masses for our populations. The blue solid line shows the median for the isolated star-forming galaxies in both panels. In both panels, the blue dash-dotted line represents the best fit to our binned isolated sources, and the grey dash-dotted line corresponds to the reference indicated in the legend. In the right panel, the blue dashed line represents the median for the star-forming interacting/merging sources. The grey points represent the isolated populations. Overall, the trend shows a mild decline in $q_\mathrm{TIR}$ with stellar mass, suggesting that more massive galaxies tend to deviate slightly from the typical infrared/radio relation.}
    \label{q-mstar}
\end{figure*}

Figure \ref{q-mstar} shows the infrared/radio correlation as a function of WISE-derived stellar masses for our objects in the isolated (left panel) and interacting/merging (right panel) samples. In both panels, the blue solid line represents the median $q_\mathrm{TIR}$ values based on WISE classification, and the blue dashed-dot line represents the best-fit to the SF populations. Indeed, we do find evidence of a decline of $q_\mathrm{TIR}$ with stellar mass: 

\begin{equation}
    q_\mathrm{TIR} ({M_\mathrm{\star}}) = (-0.128 \pm 0.031) \, \mathrm{log} \left( {M_\mathrm{\star}}/10^{10}\mathrm{M_{\odot}} \right) + 2.659 \pm 0.016
    \label{q-mstar-relation}
 \end{equation}

We found that our best-fit slope ($m =-0.128 \pm 0.031$) for the SF galaxies is very close to Delvecchio's (-0.124 $\pm$ 0.015), with the difference corresponding to only 0.1$\sigma$. Although our best-fit relation shows a higher intercept than that of \cite{Delvecchio2021}, this difference may simply reflect variations in the properties of the samples. Their relation lies within the uncertainties of our moving-average $q_\mathrm{TIR}(M_{\star})$ (see Figure~\ref{q-mstar}), indicating that the two relations are consistent within our measurement errors. This suggests that the rate of decline of $q_\mathrm{TIR}$ with stellar mass observed in our sample is similar to that found by \cite{Delvecchio2021}. Thus, our results confirm the findings of \cite{Delvecchio2021} and further demonstrate that even in the local Universe $q_\mathrm{TIR}$ exhibits a clear relation with stellar mass.

We do not attempt to fit the interacting/merging sample due to its smaller size and large dispersion, which makes it sensitive to statistical fluctuations. However, we include it for comparison with the isolated population. In particular, we investigate the variation in the $q_\mathrm{TIR}$ with our sources binned according to these derived stellar masses. Indeed, a KS test revealed that the difference in the infrared/radio correlation between the interacting/merging and isolated populations of similar stellar masses is statistically significant (statistic = 0.329, $p$-value = 0.018). Also refer to Table \ref{binned_phys_props} for statistics of the $q_\mathrm{TIR}$ across our 3 stellar mass bins.

\subsection{Outliers and other interesting sources} \label{outies}
A few objects deviate from the well-known infrared-radio relation by having excess emission in either the infrared or radio band. These objects are identified as outliers because they exhibit deviations of more than 3$\sigma$ on the $L_{1.4 \, \rm{GHz}}$ and $L_\mathrm{TIR}$ relation plot for both the total and isolated samples. Additionally, we included sources with the faintest luminosity in the sample, as well as those with the highest/lowest $q$-values from the "interacting" sample. We discuss these outliers in detail below, arranged in ascending order of $q_\mathrm{TIR}$ value. First, we start with sources in the total/Isolated samples, then finally finish with those in the merging/interacting sample. 

\subsubsection{Sources in the total/isolated samples}

NGC 1275 - This is a nearby ($z_\mathrm{cmb} = 0.01544$) central giant elliptical galaxy located in the centre of the Perseus cluster, harbouring a Seyfert $\rm 2$ nucleus that accretes gas from the x-ray emitting intracluster gas through cooling flow. It also exhibits an extended radio structure spanning $ \rm 10'$, resembling an asymmetrical Fanaroff $\rm - $ Riley type I source, with its jet axis closely aligned with the line-of-sight \citep{Aleksi2014A&A...564A...5A}. This is also believed to have undergone a merger with another galaxy a few hundred million years ago \citep{Holtzman1992AJ....103..691H, Conselice2001AJ....122.2281C}. This source exhibits the lowest $q_\mathrm{TIR}$ value in our sample, $q_\mathrm{TIR} = 0.05.$ 

NGC 5128 (Centaurus/Cen A) - This is a nearby ($z_\mathrm{cmb} = 0.00093$) giant elliptical (early-type) galaxy that hosts a strong and active radio source. It is a classical radio galaxy powered by a supermassive black hole ($M_\mathrm{BH} = 5.5 \times 10^7 \,\rm{M_\mathrm{\odot}}$; \citealt{Cappellari2009}), generating X-ray and radio jets, as well as extensive radio lobes and optical filaments in the outer parts of the galaxy. It is classified as a prototype of a low luminosity Fanaroff-Riley (FR) class I radio galaxy. Extensive research was conducted on this galaxy in a wide range of wavelengths, revealing the presence of a warped disk and optical shells with associated neutral hydrogen and CO emission; likely believed to be the aftereffects of past merger activity (\citealt{Israel1998A&ARv...8..237I}). Refer to Figure \ref{outies_images}, first panel on the left, to view our MeerKAT and WISE RGB image of this source. Its $q_\mathrm{TIR}$ value is equal to 0.38.

NGC 5793 - This is a nearby ($z_\mathrm{cmb} = 0.01233$) edge-on disk galaxy, hosting a bright compact nucleus seen in the radio continuum (see Figure \ref{outies_images}, first panel on the right). Optical emission line diagnostics classify the nucleus of this source as a Seyfert 2 \citep{Baan1998}. From its excess FIR emission ($L_\mathrm{FIR} > 10^{10.8}  L_\mathrm{\odot}$) this object was identified as being a starburst galaxy \citep{soifer1987iras}. This galaxy exhibits peculiar $\rm H_{2}O$ maser features at velocities $\rm 250 \, km/s$ around its central region, which is thought of as an indication of the presence of a rotating disk/torus \citep{Hagiwara1997}. In addition, single-dish observations of this galaxy reveal strong HI and OH absorptions towards the central compact radio nucleus \citep{Gardner1986}. This galaxy is also known to contain a CO ($\rm J = 1-0$) emission that extends up to a radius $\rm 1.2$ kpc from the centre \citep{Hagiwara1997}.

IC 5063 - This is a nearby ($z_\mathrm{cmb} = 0.01093$) elliptical galaxy with large-scale dust lanes possibly resulting from a merger \citep{Colina1991ApJ...370..102C, Morganti1998}. It is one of the most radio-loud  ($L_\mathrm{1.4 \, GHz} = 3 \times  10^{23} \, \rm{W\,Hz^{-1}}$) type 2 Seyfert galaxies found in the nearby universe \citep{Travascio2021}. This galaxy is also HI-rich. In our radio image for this source (see Figure \ref{outies_images}, second panel left), there is a short jet extending outward.

NGC 5506 - A nearby ($z \sim 0.00705$) edge-on (highly inclined) spiral galaxy, classified as Sa pec, located in the local Virgo supercluster. It hosts a Seyfert 1.9 nucleus \citep{Riffel2021} and exhibits a nuclear molecular ring-like morphology with a radial extent $\sim 50$ pc. Additionally, it contains molecular gas, as indicated by CO emission at the AGN, and significant HCO$^+$ emission stemming from very dense molecular gas in the torus \citep{Garci_Burillo2021A&A...652A..98G}. Refer to Figure \ref{outies_images}, second panel on the right, to view our MeerKAT and WISE RGB image of this source.

NGC 4151 - A type 1.5 Seyfert galaxy at $z \sim 0.00414$. Acknowledged as one of the brightest AGNs in the x-ray sky \citep{wang2010}, it also holds the title of being the radio-brightest of the radio-quiet AGNs \citep{Zdziarski2000ApJ}. The notable variability displayed by this AGN has positioned it as one of the extensively studied nearby AGNs across the electromagnetic spectrum (see \cite{Chen2023MNRAS.520.1807C} and references therein).

NGC 6822 (DDO209) - This is a nearby ($z_\mathrm{cmb}\sim 0.0012$) metal-poor ($\sim 0.2\, \rm{Z_\mathrm{\odot}}$; \citealt{Garc2016}) isolated irregular quiescent dwarf galaxy located in the Local Group. It comprises different populations with different distributions and kinematics. Its interesting characteristics (i.e., proximity, low metallicity and lack of known close companions) have made it an ideal subject of intense study. In the optical, it is characterised by a central bar oriented in the north-south direction. It encompasses most of the young stellar content of the galaxy, a huge HI disk and several prominent HII regions and OB associations \citep{Lenki2023}. Its HII regions are among the most massive and brightest known in the local universe and span various evolutionary stages \citep{Efremova2011, Jones2019}. While these thermal regions likely account for most of its radio emission, the galaxy remains exceptionally weak in the nonthermal radio continuum \citep{Condon1987, Cannon2006ApJ...652.1170C}. Refer to Figure \ref{outies_images}, third panel on the left, to view our MeerKAT and WISE RGB image of this source. Notably, it does not exhibit any extended emission in the radio continuum but instead appears rather patchy. Its radio measurements may be somewhat uncertain \citep{Condon2021}. The primary challenge lies not in baseline limitations or low surface brightness but in determining whether each detected radio component belongs to NGC 6822 or a background galaxy. Nonetheless, this source has the lowest infrared and radio spectral luminosities in the entire sample, and its high $q_\mathrm{TIR}$-value (i.e., 3.07) places it well above the nominal infrared-radio correlation. Due to its location as a Local Group member attached to our Galaxy, it is not representative of the typical space density of galaxies. From this source onward, the objects exhibit higher-than-typical $q_\mathrm{TIR}$ values.

IC 1953 - This is a late-type barred Sd galaxy at $z \sim 0.00581$. The radio continuum morphology comprises diffuse emission from the disk and an unusually bright and well-defined core for its late-type (highly prominent in the mid-infrared dust emission;  \citealt{Roussel2003}). Additionally, the galaxy is considered to be radio-deficient, with the observed synchrotron component attributed solely to the spiral disk and lacking in the central regions. After examining several scenarios, \cite{Roussel2003} suggests that this galaxy has been observed within a few Myr of the onset of an intense star formation episode following a period of quiescence. This is the first outlier in the sample whose emission is not AGN-powered, exhibiting a $q_\mathrm{TIR} = 3.35$. Refer to Figure \ref{outies_images}, third panel on the right, to view our MeerKAT and WISE RGB image of this source.

NGC 4437 (also known as NGC 4517) - This is a nearby ($z \sim 0.00218$) edge-on spiral galaxy, forming a pair with NGC 4592. Based on the nuclear emission line
ratios, \citet{Seth2008ApJ...678..116S} classified it as a composite object, suggesting that its emission could be a mix of star formation or AGN activity. 

NGC 2403 - This is a nearby ($z \sim 0.00079$) SAB(s)cd isolated galaxy in the outskirts of the M81 group. Although it lacks a central bulge, it hosts a luminous compact nuclear star cluster \citep{Yukita2010AJ....139.1066Y}. While it is rich in H II regions, no evidence of AGN activity in its nucleus has been reported in the literature, nor do we find any in our analysis.

NGC 4418 - This is a nearby LIRG ($ L_\mathrm{TIR}=10^{11.19}\, \rm{L_\mathrm{\odot}}$) located at a $z_\mathrm{cmb} = 0.00843$. It harbours a compact, bright mid-infrared emitting core at the nucleus but emits significantly less radio emission than expected. \cite{yun2001} found only $\rm 10$ such objects out of a sample containing $\rm 1809$, making NGC 4418 unusual. Despite being well-studied (\citealt{Roche1986, Sakamoto2013, Ohyama2023}), the source of the extreme luminosity in this galaxy is still largely debated. The pronounced absorption at 9.7 $\rm \mu$m, and the prominent ice features at MIR are often linked to concealed heating sources such as buried young stellar objects. These features set it apart from conventional traits observed in active galactic nuclei (AGNs) and starburst galaxies. Hence, the nucleus of NGC 4418 is occasionally referred to as a compact obscured nucleus. Refer to Figure \ref{outies_images}, fourth panel on the left, to view our MeerKAT and WISE RGB image of this source. Even in our analysis, this object exhibits a higher infrared luminosity relative to its radio emission, with a corresponding $q_\mathrm{TIR}$-value of 3.44.

NGC 0925 - This late-type spiral galaxy (SAB(s)d) at $z \sim 0.00236$ hosts an unusually high number of ultra-luminous x-ray sources \citep{Salvaggio2022MNRAS, maritza2021ApJ}. Its HI content has been extensively studied (e.g., \citealt{Pisano1998}), and \citet{Pisano2000AJ....120..763P} found evidence of widespread star formation occurring outside of the radius where it should be suppressed.

IRAS F08572+3915 - A nearby ($z \sim 0.05906$) warm ULIRG (log($L_\mathrm{TIR}) = 12.16\, {L_\mathrm{\odot}}$) with strong mid-IR silicate absorption \citep{Herrera-Camus2020}. It is composed of a pair of interacting/merging galaxies with a nuclear separation of about $\rm 6 \, kpc$ \citep{Veilleux2002}. This source is considered to be powered by an AGN, although it also exhibits strong signs of significant starburst activity, probably due to the merger. Previous optical classifications of LINER or Sy 2 were based on shallow spectra that did not clearly detect the $\rm [OIII] \, 5007$ {\AA}  or $\rm H \beta$ emission lines \citep{Herrera-Camus2020}. We found its $q_\mathrm{TIR}$ value to be 3.59, which may be uncertain or anomalous due to the merger.

NGC 1377 - This is a nearby ($z_\mathrm{cmb} = 0.00557$) lenticular galaxy that exhibits no signs of disturbance. It stands out as one of the most radio-quiet and FIR-excess galaxy  \citep{Roussel2003, Roussel2006ApJ...646..841R, Aalto2016A&A...590A..73A}. Its radio synchrotron emission is deficient by at least a factor of 37 compared to that of normal galaxies. Its central region is heavily enshrouded by dust, and the origin of the main source of the abnormal FIR luminosity (and the cause of its radio deficiency) is still elusive. On the one hand, \citet{Roussel2003, Roussel2006ApJ...646..841R} proposed that it is likely a nascent starburst. On the other hand, \citet{Imanishi2006} found evidence of strong signatures of obscured AGN in this galaxy. This source is characterised by low mass, a hot dust content $ \rm (F60/F100 \sim 1.2)$ and exhibits deep mid-IR silicate absorption features \citep{Roussel2008ASPC..381..297R}. Although determining the dust temperature is challenging, a fit to the 25–100 $\mu m$ data suggests a blackbody temperature of 80 K and a diameter of 37 pc \citep{Roussel2003}. \citet{Aalto2020A&A...640A.104A, Aalto2012A&A...546A..68A} have reported finding evidence of a large concentration of molecular gas and massive molecular outflows in this object, supporting the AGN scenario. The high nuclear dust and gas obscuration in this source makes it difficult to determine the nature of the nuclear activity. This object remains faint even at the depth of our 1.28 GHz observations, but it is extremely bright in the mid-infrared (see Figure \ref{outies_images}, fourth panel on the right). Consequently, its $q_\mathrm{TIR}$ = 4.71 is highly uncertain and anomalous.
 
\subsubsection{Source(s) in the interacting/merging sample}
\begin{figure*}
    \centering
    \subfloat{{\includegraphics[width=0.48\textwidth]{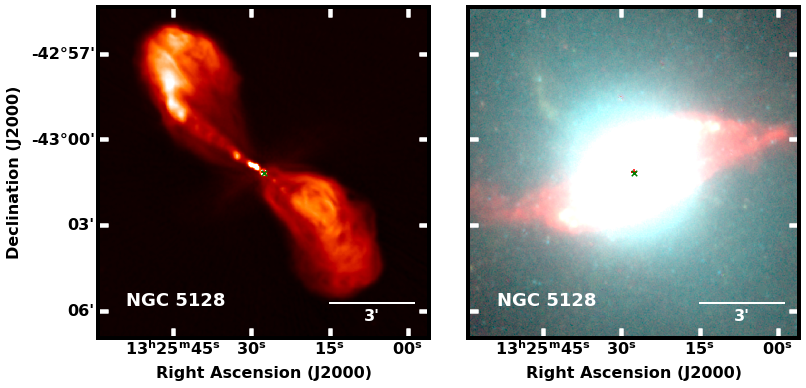} }}
    \subfloat{{\includegraphics[width=0.48\textwidth]{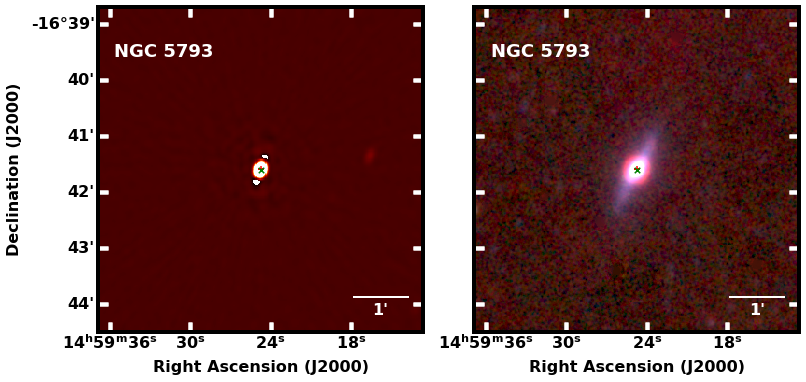} }}%
    
    \vspace{0.1pt}
    \subfloat{{\includegraphics[width=0.48\textwidth]{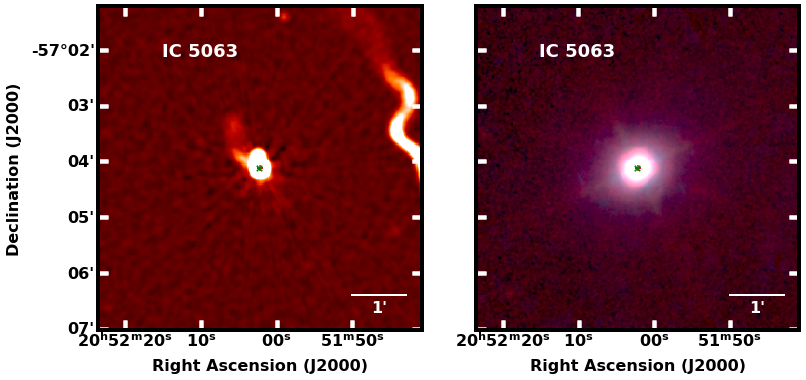} }} 
    \subfloat{{\includegraphics[width=0.48\textwidth]{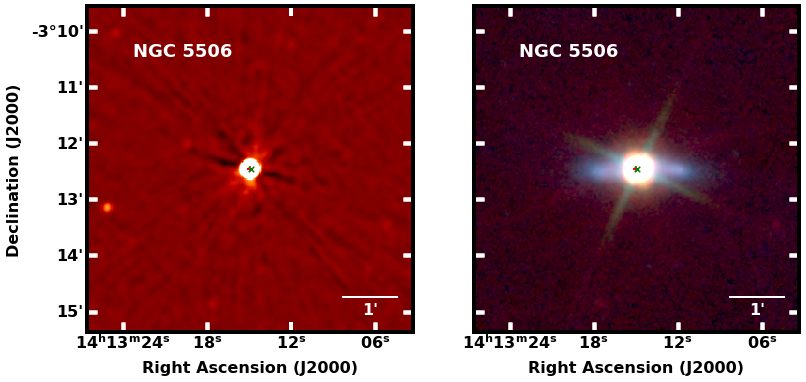} }}%
    
    \vspace{0.1pt}
     \subfloat{{\includegraphics[width=0.48\textwidth]{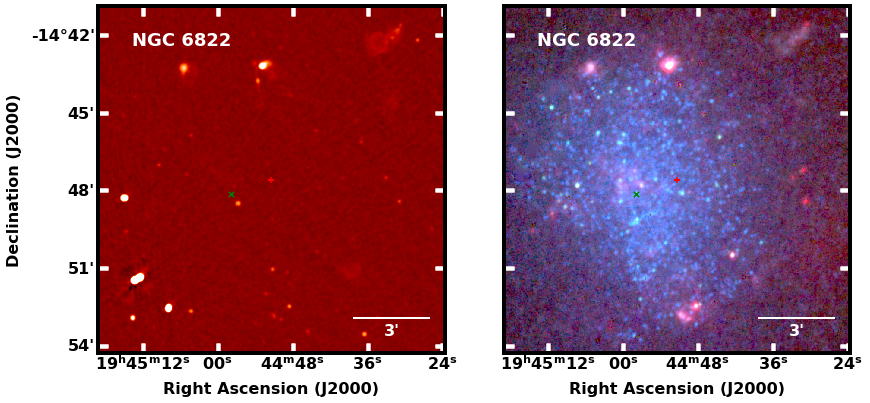} }}%
     \subfloat{{\includegraphics[width=0.48\textwidth]{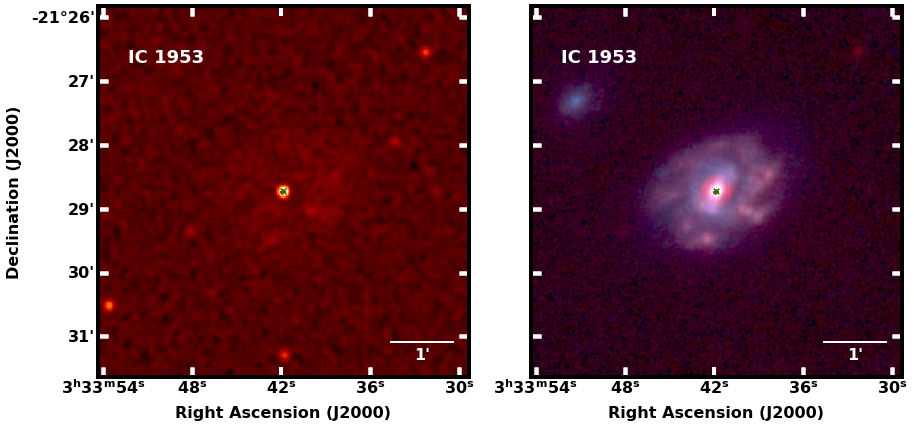} }}%
      
    \vspace{0.1pt}
    \subfloat{{\includegraphics[width=0.48\textwidth]{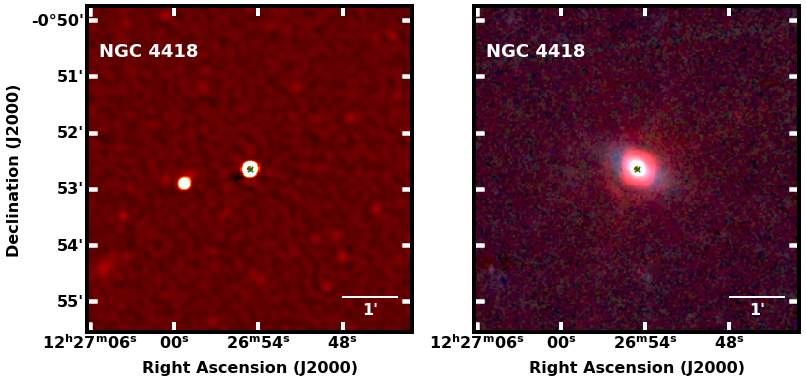} }}
    \subfloat{{\includegraphics[width=0.48\textwidth]{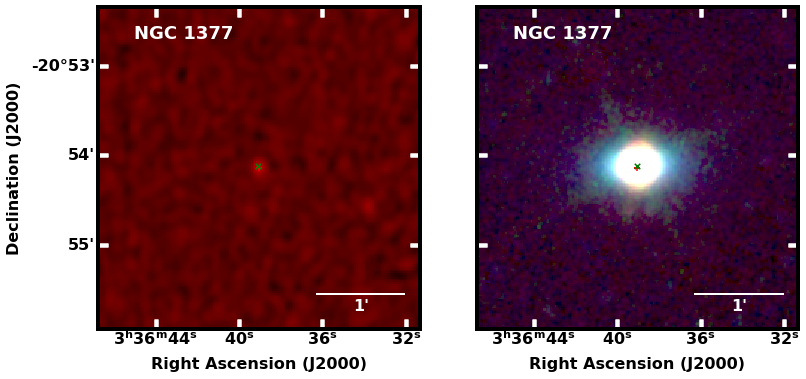} }}%
    
    \vspace{0.1pt}
    \subfloat{{\includegraphics[width=0.48\textwidth]{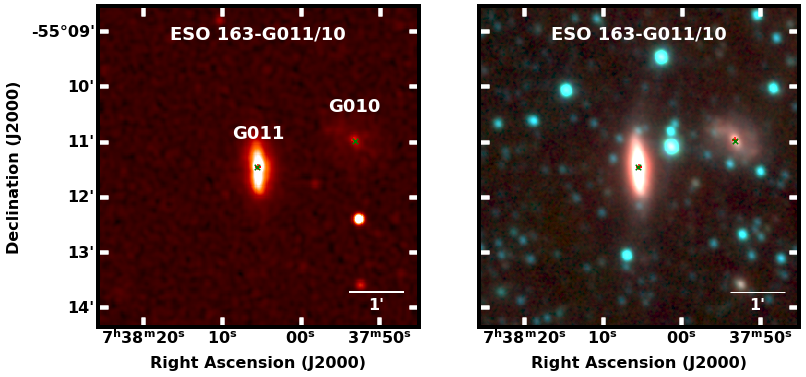} }}
    
    \caption{MeerKAT and WISE 3-colour images of some of the objects discussed in Section \ref{outies}, highlighting a portion of the sources deviating from the correlation in the total and isolated sample, as well as a source exhibiting the lowest infrared and radio spectral luminosity in the interacting/merging sample. Refer to \citet{Condon1990, Condon1996} for the corresponding radio (VLA) images of the northern RBGS sources that deviate from the correlation.}
    \label{outies_images}
\end{figure*}

ESO 163-G010/011 – There is little information available about this system in the literature. We find that both components are dominated by star formation based on our WISE colour–colour diagnostic. Notably, one of its components, ESO 163-G010, has the lowest infrared and radio luminosities in our merging/interacting sample. Refer to Figure \ref{outies_images}, last panel on the right, to view our MeerKAT and WISE RGB image of this system.

Overall, we note that deviations in some instances arise due to sources being radio-quiet, as illustrated in Figure \ref{outies_images} (fourth panel on the right; NGC 1377), while simultaneously exhibiting significant infrared activity. This contributes to substantial differences in their infrared/radio spectral luminosities. However, the majority of sources showing deviations from the infrared/radio correlation are those hosting AGN, aligning with expectations, as the excess radio emission from the AGN contributes to the overall observed emission. Additionally, we note that some of the outliers (3 sources to be specific), particularly those with lower $q_\mathrm{TIR}$ values, have likely undergone a merging process in their past.

\section{Discussion}\label{discussion}

As expected, the IRAS RBGS sample follows a tight infrared/radio correlation. For both the total and isolated samples, we found that our $\left<q_\mathrm{TIR} \right>$ results ($\rm 2.63 \pm 0.01$ and $\rm 2.61 \pm 0.01$ respectively) are consistent with other studies of the correlation conducted in the local universe (\citealt{wang2019, bell2003}) and using radio data observed with the same instrument (\citealt{yao2022}, i.e., MeerKAT). However, we observed that interacting/merging galaxies do not follow the correlation as tightly and show a larger dispersion (NMAD = 0.27) compared to isolated galaxies, which have NMAD = 0.14. Below, we discuss the interesting points drawn from our analysis. 

\subsection{The non-linear nature of the infrared/radio correlation}\label{nonlin-nature}

Understanding the linear nature (or lack thereof) of the infrared-radio correlation is essential, as it directly impacts the use of radio emission as a reliable tracer of star formation rates. For both the total and isolated samples, we found the logarithmic slope of the correlation to be close to unity ($\rm \sim 1.02$ for the total sample and $\rm \sim 1.04$ for the isolated sample), indicating a nearly proportional relationship between radio and infrared luminosities. However, in Section \ref{lum_cal_explained}, we noted a non-linear relationship between $L_\mathrm{TIR}$ and the MIR star-formation rate indicator $L_\mathrm{12\, \mu m}$ (as shown in Equation \ref{cluver_LTIR_proxy}), which we used to analyse our "interacting/merging" sample. The logarithmic TIR/MIR slope aligns with the TIR/radio slope, and their relationship is statistically significant at a $6 \sigma$ level. 

This common behaviour between TIR/radio and TIR/MIR ratios possibly attests to the non-linear nature of the correlation. For example, cosmic-ray leakage from small low-luminosity galaxies may increase the TIR/radio ratio but would not affect the TIR/MIR ratio, highlighting the complexity of these correlations. Additionally, variations in dust temperature could affect the TIR/MIR ratio without significantly impacting the TIR/radio ratio, offering another layer of complexity in understanding these relationships. These differences in how these processes influence the ratios suggest that other physical factors are at play. Previous studies have shown that metallicity ($Z$) varies with luminosity (e.g., \citealt{Zahid2011ApJ...730..137Z, Garnett2002ApJ...581.1019G}), and so does the dust temperature (e.g., \citealt{Hwang2010MNRAS.409...75H, Magnelli2014A&A...561A..86M}). The underlying causes of the luminosity variation may differ for TIR/radio and TIR/MIR ratios. Overall, the consistent slope across both the pure TIR/MIR and TIR/radio ratios provides evidence of a strong and non-linear characteristic of the correlation.

These results suggest that other factors beyond star-formation activity are at play in these objects. For instance, \citet{Delvecchio2021} examined the effect of galaxy stellar mass on the IR/radio correlation and found that the correlation evolves primarily with mass, with more massive galaxies displaying a systematically lower $q_\mathrm{TIR}$ for SF galaxies. Indeed, even in our redshift-limited sample, we find evidence of the dependence of the infrared/radio correlation on stellar mass. This supports the finding of \citet{Delvecchio2021} and our relations are consistent within measurement uncertainties (see Figure~\ref{q-mstar}), indicating that the agreement is both statistically meaningful and robust. Thus, the infrared/radio correlation, although fundamentally linked to star-formation activity, is modulated and shaped by multiple other physical processes, which leads to its non-linear nature.

\subsection{Infrared/radio correlation versus luminosity}
Understanding the luminosity dependence of $q_\mathrm{TIR}$ is crucial for interpreting its evolution and intrinsic properties. Observational evidence (e.g., \citealt{Schreiber2018A&A...609A..30S, Viero2022MNRAS.516L..30V}) has shown that dust temperature increases with redshift in star-forming galaxies. However, disentangling redshift and luminosity effects remains a challenge, as dust temperature also correlates with infrared luminosity (e.g., \citealt{Hwang2010MNRAS.409...75H, Magnelli2014A&A...561A..86M}), though the fundamental nature of this correlation remains debated (e.g., \citealt{Schreiber2018A&A...609A..30S}). Cosmic ray leakage (discussed in Section \ref{nonlin-nature}) could alter synchrotron emission. As such, any significant variations or uncertainties in infrared or radio properties could impact the infrared/radio ratio. 

By definition (Equation \ref{q_equation}), if $q_\mathrm{TIR}$ does not vary with FIR luminosity and has a non-zero dispersion,  a decrease with increasing radio luminosity is expected (see, e.g., \citealt{Condon1984ApJ...287..461C, Moric2010ApJ...724..779M}). In our sample, we find no evidence of a significant variation in $q_\mathrm{TIR}$ with infrared luminosity (not shown). We also do not observe extreme excesses or deficits in IR emission that could significantly influence $q_{\mathrm{TIR}}$. The only notable exception is NGC 1377, radio-quiet yet IR-bright, positioning it as an outlier in the nominal infrared-radio correlation. We do observe a decline of $q_\mathrm{TIR}$ with increasing radio spectral luminosity (Figure \ref{rolling_mean_com}, beyond $\rm{log} (L_{1.4\,\rm{GHz}} \rm{W\,Hz^{-1}} )>21$). This decrease in $q_\mathrm{TIR}$ with increasing radio power is consistent with previous studies (e.g., \citealt{molnar2021, Matthews2021, yao2022, Giulietti2022MNRAS.511.1408G}). For example, \citet{yao2022}, using the MeerKAT-GAMA-WISE sample, found a decrease in $q_\mathrm{TIR}$ with increasing radio spectral luminosity, which they attributed to redshift evolution. However, even in our redshift-limited ($z_{\text{cmb}}$ < 0.09) sample, we already observe this trend. This suggests that while the redshift is not a major factor in the correlation observed in this sample, there seems to be a link between how strong the correlation is and the levels of radio spectral luminosity. Hence, we propose that radio spectral luminosity, rather than redshift, is likely the primary driver of this correlation.

\citet{Matthews2021}, using their 2MASX sample, also observed a decline in their $\left<q_\mathrm{FIR} \right>$ with radio spectral luminosities below $\rm{log}[\mathit{L}_\mathrm{1.4 \, GHz} (W\,Hz^{-1})] \approx 22.5$, which then flattens at higher radio luminosities. This decline in the $q$-parameter with increasing $L_\mathrm{1.4,{GHz}}$, particularly for the isolated (i.e., pure) sample, and \citet{Matthews2021}, is very strong and well illustrated in Figure \ref{rolling_mean_com}. \citet{Matthews2021} mitigated biases in their $q_{FIR}$ distribution by dividing their sample into luminosity bins and weighting the value of $q_\mathrm{FIR}$ for each source based on a $V_\mathrm{max}$ analysis, thereby yielding an unbiased volume-limited distribution of $q_\mathrm{FIR}$. Hence, the consistency of this decline between our flux-selected sample and their volume-limited sample of SFGs is validated. Our findings, illustrated in Figure \ref{qtir-z-lrad-total} lower panel and Figure \ref{rolling_mean_com}, further corroborate this declining behaviour. Above $\rm{log}[L_\mathrm{1.4 \, GHz} (W\, Hz^{-1})] \approx 22.5$, our $60\, \mathrm{\mu m}$ flux-limited RBGS sample favours luminous SFGs, and actually has better statistics compared to the $2\, \mathrm{\mu m}$ flux-limited 2MASX sample. The decrease of $q_\mathrm{TIR}$ with increasing $L_\mathrm{1.4,{GHz}}$ is consistent with a decrease of $q_\mathrm{TIR}$ with galaxy stellar mass \citep{Delvecchio2021} because $L_\mathrm{1.4,{GHz}}$ is correlated with SFR (and stellar mass). However, we note that with the NED classification, the relation turns to change at higher spectral luminosities and eventually flattens (see last bin). This behaviour confirms the observed tentative flattening in \citet{Matthews2021}, Figure 10. This flattening is likely a result of AGN  artificially boosting the radio luminosity without proportionally affecting the infrared emission (i.e., AGN contamination). In contrast, our WISE colour-colour classification for the isolated sample shows that AGNs become prevalent at higher spectral radio luminosities, and there is no evidence of flattening for this sample. 

To determine the magnitude of the decline, \citep{Moric2010ApJ...724..779M, molnar2021} generated mock data by sampling the observed  $q_\mathrm{TIR}$ distribution and calculated $L_{1.4}$ by combining these randomised  $q_\mathrm{TIR}$ values with the real $L_{\mathrm{TIR}}$ measurements. Finally, the bootstrapped  $q_\mathrm{TIR}$-$L_{1.4}$ relation was fitted. Using a similar approach, we found that the decreasing $q_\mathrm{TIR}$-$L_{1.4}$ trend in our data is consistent with the simulated data within statistical uncertainties.

This observed decline in $q_\mathrm{TIR}$ with increasing radio spectral luminosity can be interpreted within the framework of the \textit{in-situ} galaxy evolution scenario by \cite{Lapi2014ApJ...782...69L, Lapi2018ApJ...857...22L}, as described by \citealt{Giulietti2022MNRAS.511.1408G}. In short, during the early stages, when the black hole is still small and its nuclear activity is limited, radio emission is primarily attributed to star formation in the host galaxy. This results in moderate radio spectral luminosities and a typical $q$-parameter. As the black hole grows and becomes more active, AGN-driven radio emission can contribute significantly, and in some cases, dominate over star formation-related radio emission. When this happens, the excess radio emission drives the $q$-parameter to lower values, thus deviating from the standard infrared-radio ratio (see \cite{Giulietti2022MNRAS.511.1408G} who gave a similar explanation of their lower $q_\mathrm{FIR}$). 

Excess radio emission is often an effective indicator of AGNs, as SF alone typically cannot produce enough radio emission to reach such elevated luminosity levels characteristic of AGNs. With increasing redshift, we are observing more luminous sources (i.e., radio-loud), which are AGNs. We find that only a small fraction of the AGN population exhibits lower $q$-values. The majority of the AGN population and the other classes have typical $q$-values and also adhere to the correlation. Thus, even if a significant AGN contribution is probably present in these objects (adding to both the FIR and radio emission), they would not be identified with just a simple low-$q$ threshold, as usually employed for radio-loud AGN. This implies that a simple low-$q$ AGN discriminator is unreliable except for the most powerful AGNs. This is also consistent with previous observations of AGN-bearing galaxies (e.g., \citealt{Moric2010ApJ...724..779M}).

Overall, we found that $q_\mathrm{TIR}$ varies more with the radio spectral luminosity than with infrared luminosity, consistent with previous studies (e.g., \citealt{molnar2021}). This has significant implications for models aiming to understand the radio evolution of the SFR history of the universe. Models that assume $q_\mathrm{TIR}$ is independent of $L_\mathrm{1.4,{GHz}}$ may overestimate the degree of evolution, as redshift and luminosity are correlated in flux-limited samples (e.g., \citealt{Tisanik2022}). Alternatively, it can lead to findings that $q_\mathrm{TIR}$ decreases with increasing redshift (e.g., \citealt{Delhaize2017}, \citealt{Novak2017}). Since the luminosity dependence of average $q_\mathrm{TIR}$ values is linked to the non-linearity (discussed earlier in Section \ref{nonlin-nature}) of the $q_\mathrm{TIR}$ (see Equation \ref{q_equation}), it is crucial to incorporate these dependencies into models. For example, at lower radio spectral luminosities, we have higher $q$-values and vice versa. Therefore, instead of assuming a single constant $q_{\mathrm{TIR}}$ value, radio continuum-based SFR estimate models should account for these variations. Additionally, the selection frequency of the different flux-limited samples should also be considered \citep{Condon1984ApJ...287..461C}. The RBGS is flux-limited at $\lambda = 60 \,\mu m$, which favours high-$q_{TIR}$ galaxies; the \citet{Matthews2021} sample is flux-limited at $\lambda = 2 \,\mu m$  dominated by older stars which emit very little 1.4 GHz or 60 $\mu m$, and \citet{Giulietti2022MNRAS.511.1408G} selects a 1.4 GHz flux-limited sample from a deep optical image, so it favours low-$q_{TIR}$ galaxies. The selection frequency alone affects the $q_{TIR}$ of each sample. By integrating these dependencies into models, we could better capture the underlying physics of star formation and AGN activity.

\subsection{Infrared/radio correlation scatter in interacting/merging systems}
Galactic interactions are astronomically important because they can trigger starbursts (e.g., \citealt{Mihos1994ApJ...431L...9M, Mihos1996ApJ...464..641M}), fuel active galactic nuclei (\citealt{Sanders1988ApJ...325...74S, Di-Matteo2005Natur.433..604D}), and drastically alter the dynamics of galaxies (\citealt{Toomre1972ApJ...178..623T, cox2006ApJ...650..791C}). The dramatic impact of mergers on the gas flows directly affects the magnetic fields of the systems (and vice versa), eventually changing their morphology following the motion of the gas, which will be amplified by shocks and gas inflow. Changes in the magnetic field structure, on the other hand, might influence gas flows, local collapse, and the morphology as well as the SF activity \citep{Kotarba2010ApJ...716.1438K, Kotarba2011MNRAS.415.3189K, Geng2012MNRAS.419.3571G, Rodenbeck2016A&A...593A..89R}. The downside of galactic interactions or mergers is that they can introduce complications to interpretations of the radio properties of galaxies, especially when the systems are unresolved. A classic example is the `taffy systems' \citep{Condon1993AJ....105.1730C}, which refers to interacting galaxies with a strong synchrotron-emitting gas bridge between them. When unresolved, these galaxies appear to have nearly twice the radio continuum emission compared to what is expected from the far-infrared (FIR)/radio correlation. However, when resolved, it becomes clear that the excess radio emission originates from a bridge of radio continuum connecting the galaxy pairs, leading to a lower FIR-to-radio ratio \citep{Condon1993AJ....105.1730C,condon2002AJ....123.1881C}.

In our study, using MeerKAT and WISE data, we were able to resolve some interacting/merging galaxies, allowing us to tentatively probe (due to the small size of this subsample) the impact of these galactic interactions on the infrared/radio correlation. Interestingly, it seems that even interacting/merging galaxies conform to this correlation, although not as tightly. Additionally, compared to the isolated sample ($\rm 2.61 \pm 0.01$), the merging sample has lower $q_\mathrm{TIR}$-values ($\rm 2.51 \pm  0.08$) and exhibits a greater degree of dispersion. A KS test revealed that the variation in the $q_\mathrm{TIR'}$-parameter between interacting/merging and isolated populations of similar luminosity and stellar mass is marginally significant (p-values = 0.024 and 0.018, respectively).

But why do the interacting/merging systems have lower $q_\mathrm{TIR'}$-values, and larger scatter? The high dispersion in the merging sample (0.27 vs 0.14 for the isolated sample) could indicate additional factors that contribute to the variations in the $q_\mathrm{TIR'}$ ratio. Excess radio emission in these resolved pairs, for instance, may be attributed to the presence of tidal tails or radio continuum bridges, which are not associated with ongoing star formation (e.g., \citealt{Condon1993AJ....105.1730C, Murphy2013ApJ...777...58M, Donevski2015MNRAS.453..638D}). Consequently, this can disrupt the balance between infrared and radio emission, leading to deviations from the expected correlation and a lower $q_\mathrm{TIR}$ value. Therefore, other dynamic processes occurring during galactic interactions and the energetics of their interstellar medium could be responsible for the dispersion observed in the correlation.

\cite{Murphy2013ApJ...777...58M}, studying a sample of 31 local infrared-bright starburst galaxies, found that ongoing mergers—where progenitors remain separate but share a common envelope or exhibit tidal tails—show excess radio emission relative to what the FIR-radio correlation would predict. They attributed the scatter to non-star-formation-related radio emission, originating from bridges and tidal tails connecting the merging galaxies, similar to what is observed in "taffy" systems. \citet{Donevski2015MNRAS.453..638D} hypothesised that shocks arising from galactic interactions and mergers can heat the gas and dust, accelerating particles and producing a tidal cosmic ray population. They tested their hypothesis on a sample of 43 infrared-bright star-forming interacting galaxies at different merger stages, and found that the FIR–radio correlation parameter varies noticeably over different merger stages; $q$ first decreases during the early merger stage and later increases. Their findings were consistent with those of \cite{Murphy2013ApJ...777...58M}, but they highlighted a case where certain galaxies, such as NGC 5256, cannot be classified as ‘taffy’-like due to a lack of H2 gas in the bridge connecting the two galaxies. They concluded that the gas in this system is being shock-heated, leading to shock-induced synchrotron radio emission.

In addition, galaxy morphology is an excellent indicator of past or present interactions, as it is generally understood that irregular morphology often results from collisions or close encounters. Accordingly, \cite{Pavlovic2019MNRAS.489.4557P} and \cite{Pavlovic2021SerAJ.203...15P} explored how interactions may influence the evolution of the FIR–radio correlation with redshift. As the fraction of interacting and irregular galaxies increases with redshift, deviations from the established correlation may occur. They found that galaxies with irregular morphologies have lower $q_\mathrm{FIR}$ values compared to normal disk galaxies, indicating that galaxy morphology—a proxy for interactions—can indeed influence the expected FIR–radio correlation.

Our findings, enabled by MeerKAT and WISE, demonstrate that indeed interacting/merging galaxies exhibit lower $q$-values and larger scatter compared to the isolated objects. This result is consistent with the work of \citet{Lisenfeld2010A&A...524A..27L}, who found that the deviation of two interacting/merging systems (UGC $12914/5$ and UGC $813/6$) from the standard FIR-radio correlation can be explained by the acceleration of additional electrons in large-scale shock waves resulting from the gas-dynamic interaction between the two ISMs. Additionally, though these characteristics are consistent with expectations for this type of system as demonstrated by the scenarios explained above, we cannot definitively attribute these observations to specific mechanisms, such as shock heating, without further evidence. In addition, these sources are well resolved, allowing us to clearly distinguish the individual components. This rules out the possibility of misidentifying the system as a single unresolved source or the presence of a significant, bright tidal bridge connecting the galaxies. 

We find no evidence of tidal bridges in our images for sources exhibiting low $q_\mathrm{TIR}$ ratios. There are other possible explanations for the observed lower infrared/radio ratios. For instance, \citet{Bressan2002} showed that $q_\mathrm{FIR}$ depends on the evolutionary stage of the stellar populations and the star formation history of a galaxy. Their models indicate that, depending on the evolutionary phase and the e-folding time of the SFR, the relationship between infrared or radio emission and the SFR can vary by up to an order of magnitude with the age of the starburst, across a range of star formation histories. Although some of these variations partly cancel out, they predict that $q_{\mathrm{FIR}}$ varies with starburst age, with stronger variations for starbursts with shorter e-folding times. They suggest that a low $q_{\mathrm{FIR}}$ can simply result from a particular evolutionary stage, such as the post-starburst phase, during which the FIR emission declines while the longer-lived nonthermal synchrotron emission dominates, enhancing the apparent radio flux. Thus, the observed scatter in $q$ may naturally arise from the galaxies being caught at different points in a bursty star formation cycle. An alternative explanation for the lower $q_\mathrm{TIR}$ is free-free absorption lowering the 1.4 GHz flux densities of compact radio sources in U/LIRGs formed by mergers \citep{Condon1991}. 
However, we cannot entirely exclude the existence of faint or diffuse tidal bridges, as such features could fall below the detection limits of our radio observations. HII-dominated IC 4686, part of a triple system (see Section \ref{outies}), has the highest $q_\mathrm{TIR}$ in our merging/infrared sample. The slightly elevated $q_\mathrm{TIR}$ in these systems can be attributed to a starburst scenario, in which interactions among the galaxies enhance infrared emission (e.g., \citealt{Telesco1988ApJ...329..174T}), while radio synchrotron emission may lag due to cosmic-ray production timescales. Indeed, we found that the interacting/merging galaxies tend to have higher $W2-W3$ colours (often used as SFR proxy), star formation rates and specific star formation rates compared to the isolated sources. These differences are statistically significant (see Section \ref{host_props_mer_v_iso}) and support the notion that mergers can trigger SF. 
 
Further evidence of the impact of galactic interactions or mergers on the infrared-radio correlation comes from the observation of certain outlier objects in our isolated samples, particularly those with lower $q_\mathrm{TIR}$. Notably, the post-merger remnants NGC 5128 \citep{Israel1998A&ARv...8..237I}, IC 5063 \citep{Colina1991ApJ...370..102C, Morganti1998}, and NGC 1275 \citep{Holtzman1992AJ....103..691H, Conselice2001AJ....122.2281C} also exhibit greater dispersion and lower $q_\mathrm{TIR}$. These findings provide additional evidence that galactic interactions or mergers could impact the infrared-radio correlation.

In a nutshell, the large scatter in the $q_\mathrm{TIR}$ parameter of this subsample points to complexities introduced by galaxy mergers. Furthermore, the departure from the FIR-radio correlation in interacting/merging systems could have profound implications for the estimation of star-formation rates. Specifically, overestimations could occur if excess radio emission from tidal shocks is misinterpreted as a star formation signature. A larger sample of interacting/merging galaxies will be needed to further analyse and confirm these trends.
\section{Summary and Conclusions}\label{conclusion}

This study provides a complete radio counterpart catalogue of the IRAS RBGS sample. We provide updated distances for all the objects as well as updated infrared and radio properties. We make use of recent MeerKAT observations at 1.28 GHz of all the RBGS sources in the southern hemisphere and combine them with VLA observations at 1.425 and 1.49 GHz of all the RBGS sources in the northern hemisphere to probe, for the first time, the TIR/Radio correlation of the full RBGS sample. We employ both literature-based classifications and WISE-based classifications to distinguish the sources into star-forming-dominated or AGN-dominated sources and study them accordingly. Moreover, thanks to the resolving power of MeerKAT, combined with WISE observations, we have been able to resolve and study some southern RBGS interacting/merging systems in greater detail for the first time in the radio. We first present the results for the full RBGS sample as described by \cite{Sanders2003}, also referred to as "Total Sample", and then proceed to analyse the isolated and interacting/merging subsamples separately. Finally, we estimate the physical properties of the southern sample using WISE mid-infrared data and investigate how they relate to the infrared/radio correlation. Our main results are summarised as follows:

\begin{itemize}
    \item The RBGS galaxies exhibit a tight and slightly non-linear infrared/radio correlation as expected. For the total sample, which also comprises IRAS unresolved galaxies with various interaction stages, we find a median $q_\mathrm{TIR}$ of $2.63 \pm 0.01$ with scatter of 0.17 based on the source classification from the literature. When we instead only focus on the RBGS isolated galaxies, thus excluding the interacting/emerging system, and we use WISE to classify our objects, we get a value of $\left< q_\mathrm{TIR}\right> = 2.61 \pm 0.01$ with a dispersion of 0.16. For interacting/merging galaxies, we estimate the $q_\mathrm{TIR'}$ to be $2.51\pm0.08$ with scatter of 0.26. Our TIR/radio correlation results are comparable with previous work that studied this correlation. In addition, we found that sources with different morphological classes significantly overlap with each other in both $q_\mathrm{TIR} \,-\, z$ and $q_\mathrm{TIR} \,-\, \mathrm{log} [L_\mathrm{1.4\, GHz} \mathrm{(W\, Hz^{-1})} ]$ parameter spaces, and also conform to the correlation.

    \item We identified certain sources whose infrared/radio ratio exceeds expected levels, which we refer to as outliers. In some cases, even sources with AGN contamination adhere to the correlation. However, in other instances, the excess radio emission from the AGN results in deviations from the infrared/radio correlation. Additionally, we observed that some deviations arise when the objects have relatively low radio emission, affecting the balance and leading to an excess of infrared emission relative to the radio. In some cases, the objects exhibit higher infrared emission due to intrinsic properties, independent of AGN influence. Therefore, caution must be exercised when using this correlation as an AGN identifier, as some galaxies with AGN still conform to the correlation.

    \item We observe a slight decline of $q_\mathrm{TIR}$ with the increasing radio spectral luminosity for all our samples (see Figure \ref{rolling_mean_com}). This decline is more pronounced in the isolated sample, and it is consistent with findings from previous work (e.g, \citealt{Matthews2021}). This suggests that the infrared-radio correlation varies with radio luminosity.

    \item We considered the non-linear behaviour of the infrared/radio correlation by exploring the relation between $L_\mathrm{TIR}$ and the MIR star-formation rate indicator $L_\mathrm{12\, \mu m}$ utilised for our interacting/merging sample. We found a strong and consistent slope (statistically significant at the $6\sigma$ level) in the difference between the TIR/radio and TIR/MIR ratios.

    \item We found that $\left< q_\mathrm{TIR}\right>$ for the isolated sources in the total sample (classified using NED) flattens at higher radio spectral luminosity compared to the WISE classified isolated sample. This flattening could be attributed to AGN contamination in the NED-classified sample.

    \item We found that our isolated SF sources exhibit a correlation between their stellar masses and star formation rates. However, the correlation does not appear to bend or flatten toward higher stellar masses. Even the interacting/merging populations also adhere to the star formation main sequence, but they have a slight tendency to shift towards regions of higher star formation.
    
    \item Using the isolated SF population only, we confirmed that even in the nearby universe ($z<0.1$), the infrared/radio correlation appears to vary with $M_{\star}$. This supports the variation that we observed with radio spectral luminosity and is consistent with expectations from the literature (e.g., \citealt{Delvecchio2021}).
    
    \item To further investigate the impact of galactic interactions on the infrared/radio correlation, we studied a subset of interacting/merging galaxies that we could resolve with MeerKAT and WISE and compared them to the isolated population. We found that interacting/merging sources have lower $q_\mathrm{TIR}$ and show greater scatter compared to isolated objects. This supports our result that galactic interactions or mergers could impact the infrared-radio correlation, and caution is warranted, especially as their abundance increases at higher redshifts. Further evidence comes from some isolated sources with a history of interactions (e.g., IC 5063), which tend to deviate from the nominal infrared/radio correlation, exhibiting lower $ q_\mathrm{TIR}$ values. Overall, we found that the difference between $SFR_\mathrm{MIR,corr}$, W2-W3 colour, $sSFR_\mathrm{MIR,corr}$ and $q_\mathrm{TIR}$ of interacting/merging and that of isolated sample matched by luminosity or stellar mass is meaningful. 
\end{itemize}

\section*{Acknowledgements} 
We acknowledge the use of the ilifu cloud computing facility – www.ilifu.ac.za, a partnership between the University of Cape Town, the University of the Western Cape, Stellenbosch University, Sol Plaatje University and the Cape Peninsula University of Technology. The ilifu facility is supported by contributions from the Inter-University Institute for Data Intensive Astronomy (IDIA – a partnership between the University of Cape Town, the University of Pretoria and the University of the Western Cape), the Computational Biology division at UCT and the Data Intensive Research Initiative of South Africa (DIRISA). This work made use of the CARTA (Cube Analysis and Rendering Tool for Astronomy) software (DOI 10.5281/zenodo.3377984 –  https://cartavis.github.io). MEM, LM and MV acknowledge financial and technical support from the Inter-University Institute for Data Intensive Astronomy (IDIA) and from the South African Department of Science and Innovation’s National Research Foundation under the ISARP RADIOSKY2020 and RADIOMAP Joint Research Schemes (DSI-NRF Grant Number 150551) and the CPRR Projects (DSI-NRF Grant Number SRUG2204254729 and SRUG22031677). MEM acknowledge financial support from the African Astronomical Society (AfAS). This research has made use of the NASA/IPAC Extragalactic Database, which is funded by the National Aeronautics and Space Administration and operated by the California Institute of Technology. The MeerKAT telescope is operated by the South African Radio Astronomy Observatory, which is a facility of the National Research Foundation, an agency of the Department of Science and Innovation. The National Radio Astronomy Observatory is a facility of the National Science Foundation, operated under a cooperative agreement by Associated Universities, Inc. We thank the anonymous referee for a careful reading and detailed suggestions for clarifying our manuscript.

\section*{Data Availability}

The data underlying this article is available in a machine-readable format on Zenodo at \url{https://doi.org/10.5281/zenodo.16889781}. The MeerLIRGs MeerKAT raw data are available as programmes DDT-20200520-TM-01 and SCI-20210212-TJ-01 from SARAO's MeerKAT Archive at \url{https://archive.sarao.ac.za}, while the MeerLIRGs MeerKAT images have already been published in \cite{Condon2021} and are
available at \url{https://doi.org/10.48479/dnt7-6q05}.



\bibliographystyle{mnras}
\bibliography{main} 



\appendix
\section{WISE and NED SF/AGN classification}\label{wise-ned_class_perfomance}
\begin{figure}
    \centering
    \includegraphics[width=0.48\textwidth]{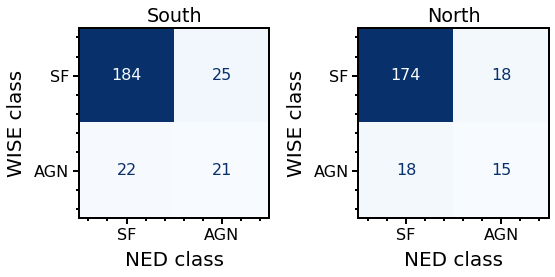}
    \caption{Confusion matrix comparing the WISE and NED SF and AGN classifications for the isolated objects in the southern and northern hemispheres, respectively.}
    \label{confusion_matrix}
\end{figure}

We evaluate the agreement between the NED and WISE AGN/SF classification results for the isolated sources using the confusion matrix formalism used in machine learning (see Figure \ref{confusion_matrix}).
We define a true positive (TP) as a source classified as an AGN by both schemes, and a true negative (TN) as a source classified as star-forming (SF) by both. A false positive (FP) refers to a source classified as an AGN by NED but as SF by WISE, while a false negative (FN) refers to a source classified as SF by NED but as an AGN by WISE.
To quantify the level of agreement, we adopt standard machine learning metrics such as accuracy, precision, recall and F1-score: we find overall values of 0.81, 0.49, 0.46, 0.47 for the southern sample and of 0.84, 0.45, 0.45, 0.45 for the northern sample, indicating a relatively good overall agreement between the two classification schemes, while highlighting some remaining inconsistencies in the identification of AGNs using the two classification schemes. We maintain that for our purposes, our careful WISE photometric analysis provides an advantage in case of disagreement.

\section{IRAS and WISE total infrared luminosity comparison for the southern isolated sample}\label{IRAS_WISE_lum_comp_appendix}
\begin{figure}
    \centering
    \includegraphics[width=0.49\textwidth]{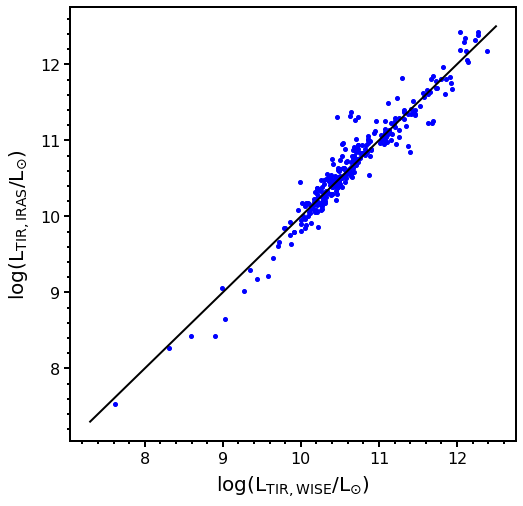}
    \caption{Infrared luminosities derived from IRAS measurements are compared to those from WISE measurements for isolated objects in the southern hemisphere (Pearson r = 0.936, p = 1.95e-129). The majority of these objects closely align with the one-to-one line, suggesting that if interacting/merging objects were resolvable by IRAS, they would exhibit a similar behaviour. Consequently, we utilise WISE luminosities as an approximation for the properties of interacting/merging objects.}
    \label{IRAS_WISE_lum_comp_appendixi}
\end{figure}

We compared the total infrared luminosities derived using IRAS and WISE measurements for the southern isolated sources. Our analysis revealed that these objects lie close to the one-to-one line (see Figure \ref{IRAS_WISE_lum_comp_appendixi}). This means that for the interacting/merging sources, which are unresolved by IRAS, we can expect a similar trend if they were resolvable by IRAS. Hence, we do not anticipate any significant bias in our derived quantities due to this effect. Consequently, we can confidently employ the WISE information to complement our IRAS results.

\section{Interacting/merging systems dominated by different mechanisms}\label{diff_mech}
We show MeerKAT and WISE 3-colour images of some of the interacting objects in which one object is powered by star formation, while the other is dominated by AGN.

\begin{figure}
    \centering
    \includegraphics[width=0.49\textwidth]{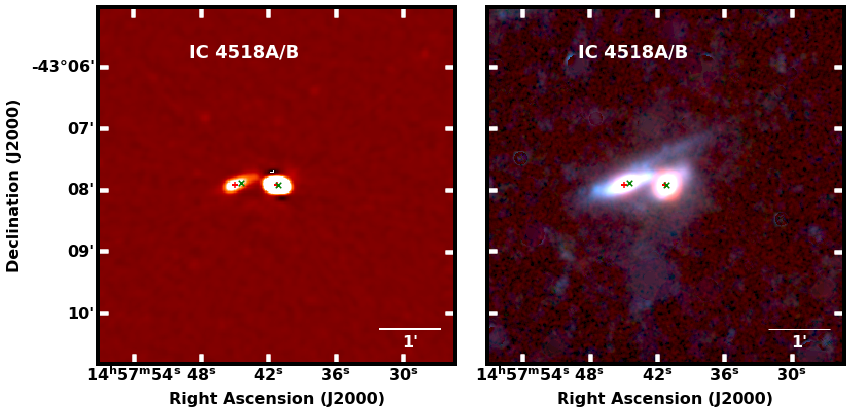}\\
    \includegraphics[width=0.49\textwidth]{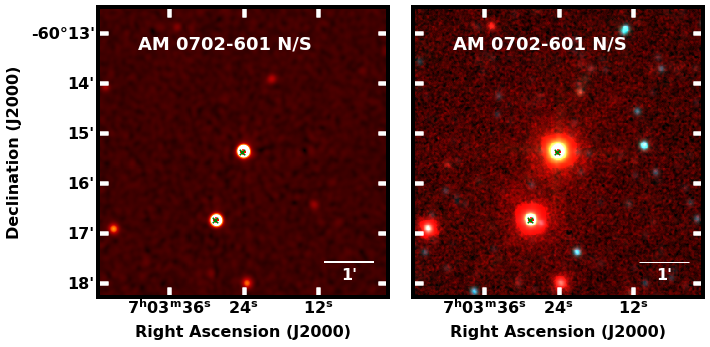}
    \caption{MeerKAT and WISE 3-colour images of interacting/merging systems with components powered by either AGN or star formation. The top panel shows the interacting/merging system IC 4518 A/B: the left component (B) is dominated by star formation, while the right component (A) is dominated by an AGN. The bottom panel shows the interacting/merging system AM 0702-601 N/S: the North (N) component is dominated by AGN emission, and the South (S) component by star formation. The red and green crosses represent the position of MeerKAT and WISE measurements, respectively.}
    \label{diff_mechanism}
\end{figure}



\bsp	
\label{lastpage}
\end{document}